\documentclass[journal]{IEEEtran}

\usepackage{subeqn}
\usepackage{graphicx}
\usepackage{float}
\usepackage{mathrsfs}
\usepackage{amsfonts}
\usepackage{cite}
\usepackage{color}
\usepackage{subfigure}
\usepackage{booktabs}
\usepackage{multirow}

\usepackage{algorithm}
\usepackage{algpseudocode}
\usepackage{graphics}

\ifCLASSINFOpdf
\else
\fi
\begin{document}

\title{Frequency-Asynchronous Multiuser Joint Channel-Parameter Estimation, CFO Compensation and Channel Decoding}
\author{Taotao Wang, and Soung Chang Liew
\thanks{The authors are with the Department of Information Engineering and the Institute of Network Coding, The Chinese University of Hong Kong, Hong Kong. Email: \{ttwang, soung\}@ie.cuhk.edu.hk. 
	
	This work is partially supported by the General Research Funds (Project No. 414713) and AoE grant E-02/08, established under the University Grant Committee of the Hong Kong Special Administrative Region, China. This work is also partially supported by the China NSFC grant (Project No. 61271277).}

}
\markboth{Frequency-Asynchronous Multiuser Joint Channel-Parameter Estimation, CFO compensation and Channel Decoding}%
{Shell \MakeLowercase{\textit{et al.}}: Bare Demo of IEEEtran.cls
for Journals}
\maketitle
\vspace{-0.5in}

\begin{abstract}
This paper investigates a channel-coded multiuser system operated with orthogonal frequency-division multiplexing (OFDM) and interleaved division multiple-access (IDMA). In general, there are many variations to multiuser systems. Our choice of the combination of OFDM and IDMA is motivated by its ability to achieve multiuser diversity gain in frequency-selective multiple-access channels. However, to realize this potential advantage of OFDM-IDMA, two challenges must be addressed. The first challenge is the estimation of multiple channel parameters. An issue is how to contain the estimation errors of the channel parameters of the multiple users, considering that the overall estimation errors may increase with the number of users because the estimations of their channel parameters are intertwined with each other. The second challenge is that the transmitters of the multiple users may be driven by different RF oscillators. The associated \emph{frequency asynchrony} may cause multiple CFOs at the receiver. Compared with a single-user receiver where the single CFO can be compensated away, a particular difficulty for a multiuser receiver is that it is not possible to compensate for all the multiple CFOs simultaneously. To tackle the two challenges, we put forth a framework to solve the problems of multiuser channel-parameter estimation, CFO compensation, and channel decoding jointly and iteratively. The framework employs the space alternating generalized expectation-maximization (SAGE) algortihm to decompose the multisuser problem into multiple single-user problems, and the expectation-conditional maximization (ECM) algorithm to tackle each of the single-user subproblems. Iterative executions of SAGE and ECM in the framework allow the two aforementioned challenges to be tackled in an optimal manner. Simulations and real experiments based on software-defined radio (SDR) indicate that, compared with other approaches, our approach can achieve significant performance gains.
\end{abstract}


\IEEEpeerreviewmaketitle

\section{Introduction}

\IEEEPARstart{T}{his} paper investigates a channel-coded multiuser system operated with orthogonal frequency-division multiplexing (OFDM) and interleaved division multiple access (IDMA). While having many advantages, a major problem of OFDM-IDMA is the \emph{frequency asynchrony} caused by the multiple carrier frequency offsets (CFOs) of the signals simultaneously transmitted by the multiple users. We put forth a framework that jointly performs multiuser channel-parameter estimation, CFO compensation and channel decoding that addresses the multiple-CFO problem in a comprehensive and systematic manner.
\\

\noindent \emph{Why IDMA?}

For a multiuser system in which multiple users transmit simultaneously to a common receiver, IDMA is a technique for facilitating the separation of user signals at the receiver \cite{ping2006interleave}. In IDMA, different transmitters interleave their channel-coded symbols in different ways before transmission to create orthogonality among user signals. In that light, it is similar to code-division multiple access (CDMA) except that CDMA creates semi-orthogonality with different spreading codes rather than different interleavers. Thus, it is not surprising that, as with CDMA, IDMA can provide multiuser diversity gains and mitigate inter-cell interference \cite{ping2006interleave, ping2005interleave}. It has been shown that, all things being equal, IDMA outperforms CDMA in terms of error rate and receiver complexity \cite{kusume2012idma}. 
\\

\noindent \emph {Why IDMA with OFDM?}

This paper focuses on wideband communications. The channels in wideband communications are often frequency-selective because of  multiple channel paths. The direct application of IDMA to wideband communications using time-domain signals leads to highly complex multiuser detectors \cite{ping2006interleave}. This is because the multiuser detectors must deal with multiple-access interference (induced by multiple user signals) and inter-symbol interference (induced by multiple channel paths) at the same time. 

OFDM is a multicarrier modulation technique that can combat frequency selectivity in wideband channels. Specifically, OFDM divides a user data stream into many parallel sub-streams and transmits them over compactly spaced subcarriers, thereby converting a frequency-selective channel into a group of frequency-flat sub-channels \cite{li2006orthogonal}. It is desirable to combine IDMA with OFDM, so that the multiuser detectors operating in the frequency domain only have to attend to multiple-access interference rather than both multiple-access interference and inter-symbol interference \cite{ping2007ofdmidma}. 

A no less important advantage of OFDM for multiuser systems is that it can tolerate unaligned symbol arrival times among the multiple user signals at the receiver. Specifically, as long as the symbol arrival times of different user signals are within the cyclic prefix (CP) of each other, the signal samples in the frequency domain are automatically aligned after the DFT \cite{lu2012implementation}. 
\\

\noindent \emph {Major Challenges in OFDM-IDMA }

Despite its advantages, the OFDM-IDMA system is susceptible to multiple CFOs among the signals of multiple users. The multiple CFOs are caused by the different RF oscillators used at the different transmitters. CFO causes inter-carrier interference (ICI) among different subcarriers and induces cumulative phase drifts over the data frame \cite{speth1999optimum}.   

Before the negative effects of CFOs  can be alleviated, an issue is the estimation of channel parameters, which include the CFOs as well as the channel gains of different users. For multiuser OFDM-IDMA, the overall estimation errors may increase with the number of users. How to contain the estimation errors in OFDM-IDMA is a major challenge.  

With accurate estimates of CFOs, the next issue is to alleviate the negative effects of CFOs. A possibility is to attempt to compensate for them at the receiver. For single-user OFDM systems, the receiver can compensate for the single CFO by multiplying the time-domain signal (before DFT) with the complex exponent of the inverse CFO. This inverse operation cancels out the CFO. After that, the receiver performs standard channel decoding in the frequency domain to extract the source message.

This method of separating CFO compensation and frequency-domain channel decoding, however, does not work for multiuser OFDM systems because of the multiple CFOs. Fundamentally, compensating for these CFOs simultaneously is impossible even if the CFOs were perfectly estimated without errors. This is because removing one of the CFOs in the received signal will necessarily leave behind some residual CFOs for the other CFOs, unless the CFOs of different users were exactly the same to begin with. As a consequence, the CFO-induced ICI inevitably remains in the frequency domain. Since the effects of all CFOs cannot be eliminated in one shot, we need an iterative method for multiuser joint CFO compensation and channel decoding for OFDM-IDMA systems.  
\\

\noindent \emph {Our Approaches and Solutions}

We put forth a framework that jointly performs multiuser channel-parameter estimation, CFO compensation and channel decoding in an integrated manner.  Our framework combines the space alternating generalized expectation-maximization (SAGE) \cite{sage1994} and  expectation-conditional maximization (ECM) \cite{mclachlan2007algorithm} algorithms.  As far as we know, this is the first attempt to construct such a unified framework for OFDM-IDMA systems. Although SAGE and ECM algorithms are well known, many gaps, however, still need to be filled in order to build a complete and consistent framework with superior performance for multiuser joint channel-parameter estimation, CFO compensation and channel decoding. This paper fills such gaps, as elaborated below. 

The framework first employs the SAGE algorithm to decompose the multisuser problem into multiple single-user problems, and iterates among single-user problems. Then the framework employs the ECM algorithm to tackle each of the single-user subproblems. Iterative executions of SAGE and ECM in the framework allow the two aforementioned challenges to be tackled in an optimal manner. 

After SAGE decomposes our multiuser problem into multiple single-user problems, ECM solves the joint channel-paramter estimation, CFO compensation and channel decoding problem for each user. Here, a key element of our framework is to exploit the sum-product message passing algorithm \cite{wiberg1996codes, kschischang2001factor} for channel decoding to refine the channel-parameter estimation as well as the CFO compensation in an iterative manner. The sum-product channel decoding corresponds to treating the coded symbols as the hidden data in the ECM algorithm. We can also adopt anthoer message passing algorithm --- the min-sum message passing algorithm \cite{wiberg1996codes, kschischang2001factor} --- to perform channel decoding. We will explain in Section III.D that the min-sum channel decoding corresponds to a  pure SAGE framework (rather than the SAGE-ECM framework). We will further show in Section IV of this paper that the proposed SAGE-ECM framework with the sum-product algorithm has better performance than the SAGE framework with the min-sum algorithm. 

To apply the sum-product algorithm in our framework, a subtle issue must be addressed.  Specifically, the sum-product algorithm, when applied for channel decoding, yields soft information on the data symbols in the frequency domain. In our iterative framework, this data soft information will in turn be used to refine channel-parameter estimation. But the CFO estimation is only feasible in the time domain (this will be elaborated in Section III.C). Therefore, to use the data for CFO estimation, we need to first transform the data soft information from the frequency domain to the time domain. This problem can be treated as a soft IDFT problem: i.e., performing IDFT on probability functions. As will be elaborated, exact soft IDFT computation can be highly complex (of exponential order). To reduce complexity in soft IDFT, we adopt Gaussian message passing \cite{loeliger2007factor} to obtain approximate solutions. We show in Section III.C that Gaussian message passing reduces the complexity from exponential order to linear order. 

In our simulations, we show that our joint framework has around 5---8 dB SNR gain over conventional multiuser approaches for systems with 2---3 users. We further performed real experiments using software-defined radio (SDR) to verify our approach. The experimental and simulated results are consistent with each other.

\subsection{ Related Works }

To our best knowledge, there is no such work that considers the channel-paramter estimation, CFO compensation and channel decoding togather for multiuser OFDM systems.

Multiuser decoding in IDMA is a method that jointly performs multiple-access interference cancellation and channel decoding \cite{ping2006interleave, ping2005interleave}. Ref. \cite{ping2007ofdmidma} proposed the use of OFDM-IDMA for the multiuser communication. However, \cite{ping2007ofdmidma} assumed the absence of CFOs,  and it directly applied the multiuser decoding technique originally developed for IDMA \cite{ping2006interleave, ping2005interleave} to OFDM-IDMA to deal with multiple-access interference in the frequency domain. Without CFOs, the application of IDMA in the frequency domain is essentially the same as that in the time domain without multipath. As mentioned earlier, unavoidable multiple CFOs can cause both inter-carrier interference among subcarriers of the same user and among subcarriers of different users. Our work here investigates this fundamental yet practical issue. 

Subsequent to \cite{ping2007ofdmidma}, \cite{peng2012improved} and \cite{dang2012experimental} considered CFOs in OFDM-IDMA. The methods of \cite{peng2012improved, dang2012experimental} cancel both inter-carrier interference (induced by CFOs) and multiple-access interference, and iterates between interference cancellation and channel decoding. However, perfect knowledge of CFOs and channel gains were assumed with no consideration given to their estimation. The methods of \cite{peng2012improved, dang2012experimental} cancel both inter-carrier interference (induced by CFOs) and multiple-access interference, and iterates between interference cancellation and channel decoding. Our simulation results in Section IV indicate that imperfect channel-parameter estimation using preambles and pilots may cause significant performance penalties (more than 10 dB) to the system. This motivates us to improve the accuracy of channel-parameter estimation for OFDM-IDMA using not just the preambles and pilots, but also the data symbols in the signals. Doing so requires iterations between the channel-parameter estimation in the time domain and channel decoding in the frequency domain. 

An alternative multiuser scheme to OFDM-IDMA is orthogonal frequency-division multiple access (OFDMA) uplink.\footnote {We remark that in the OFDM-IDMA system of interest to us, all the subcarriers are used by all users. This is different from OFDMA, where different users use non-overlapping subcarriers. Multiple-access in OFDM-IDMA is achieved by means of user-specific interleaving in IDMA. OFDM-IDMA has better spectrum efficiency than OFDMA.} As in OFDM-IDMA, the multiple CFOs in the OFDMA uplink cannot be compensated for in one shot at the receiver. Thus, in the presence of multiple CFOs, the user signals of OFDMA uplink will overlap in the frequency domain (i.e., these subcarriers are not strictly orthogonal due to the CFOs). The authors of \cite{huang2005interference, manohar2007cancellation} derived multiuser decoding methods for OFDMA to cancel the multiple-user interferences (induced by CFOs) and decode the data symbols in an iterative manner. As with the investigations of OFDM-IDMA in \cite{peng2012improved, dang2012experimental}, the investigations of OFDMA in \cite{huang2005interference, manohar2007cancellation} did not consider the impact of imperfect CFO and channel estimations. 

Recently, \cite{pun2007iterative} proposed a method for multiuser joint channel-parameter estimation, CFO compensation and symbol detection for OFDMA. As with our current work, the method of \cite{pun2007iterative} also aims to combine SAGE and ECM for solving the problem. However, \cite{pun2007iterative} did not consider the impact of channel coding, which necessitates a total recast of the algorithmic framework. In this work, we explore the proper way to combine SAGE and ECM for channel-coded OFDM-IDMA. The method of \cite{pun2007iterative} can also be modified for application to channel-coded OFDM-IDMA in a straightforward manner. We will show that a simple extension of the method of \cite{pun2007iterative} to the channel-coded OFDM-IDMA leads to a worse performance than our method. 

The rest of this chapter is organized as follows. Section II describes our system model. Section III presents our framework and shows how to apply  SAGE and ECM to OFDM-IDMA for multiuser joint channel-parameter estimation and channel decoding. Section IV details our  the simulation and experimental results. Section V concludes this paper.

\textbf{Notations}: We denote matrices by bold capital letters, vectors by bold small letters, and scalars by regular letters throughout this chapter. All vectors are column vectors. The ${\left( {i,j} \right)^{th}}$ entry of matrix ${\bf{A}}$ is denoted by ${\left[ {\bf{A}} \right]_{i,j}}$. In addition, ${{\bf{A}}^{\rm{T}}}$, ${{\bf{A}}^{\rm{H}}}$, ${{\bf{A}}^{ - 1}}$ and $\det \left( {\bf{A}} \right)$ denote the transpose, the conjugate transpose, the inverse and the determinant of ${\bf{A}}$, respectively. ${\mathop{\rm Re}\nolimits} \left(  \cdot  \right)$ means the real part of one complex number, and  $\angle \left(  \cdot  \right)$ is the angle of a complex number. ${\cal C}{\cal N}\left( {{\bf{x}}:{\bf{m}},{\bf{K}}} \right) \buildrel \Delta \over = \frac{1}{{{\pi ^r}\det \left( {\bf{K}} \right)}}\exp \left[ { - \left( {{\bf{x}} - {\bf{m}}} \right)^{\rm{H}}{{\bf{K}}^{ - 1}}{{\left( {{\bf{x}} - {\bf{m}}} \right)}}} \right]$ denotes the probability density function (PDF) of an $r$-dimension complex Gaussian random variable ${\bf{x}}$  with mean vector ${\bf{m}}$  and covariance matrix ${\bf{K}}$. The Euclidean norm of a vector ${\bf{x}}$ is denoted by $\left\| {\bf{x}} \right\|$. Finally, $\oplus $ denotes the exclusive-or operation.

\section{System Model}


\subsection{Transmit Signal}

Let us first look at the transmit side of our system model. In the uplink, $U$ users transmit simultaneously to a base station. The transmitted messages employ OFDM signaling  with IDMA. 

\begin{figure*}[!t]
	\centering
	\includegraphics[width=6in]{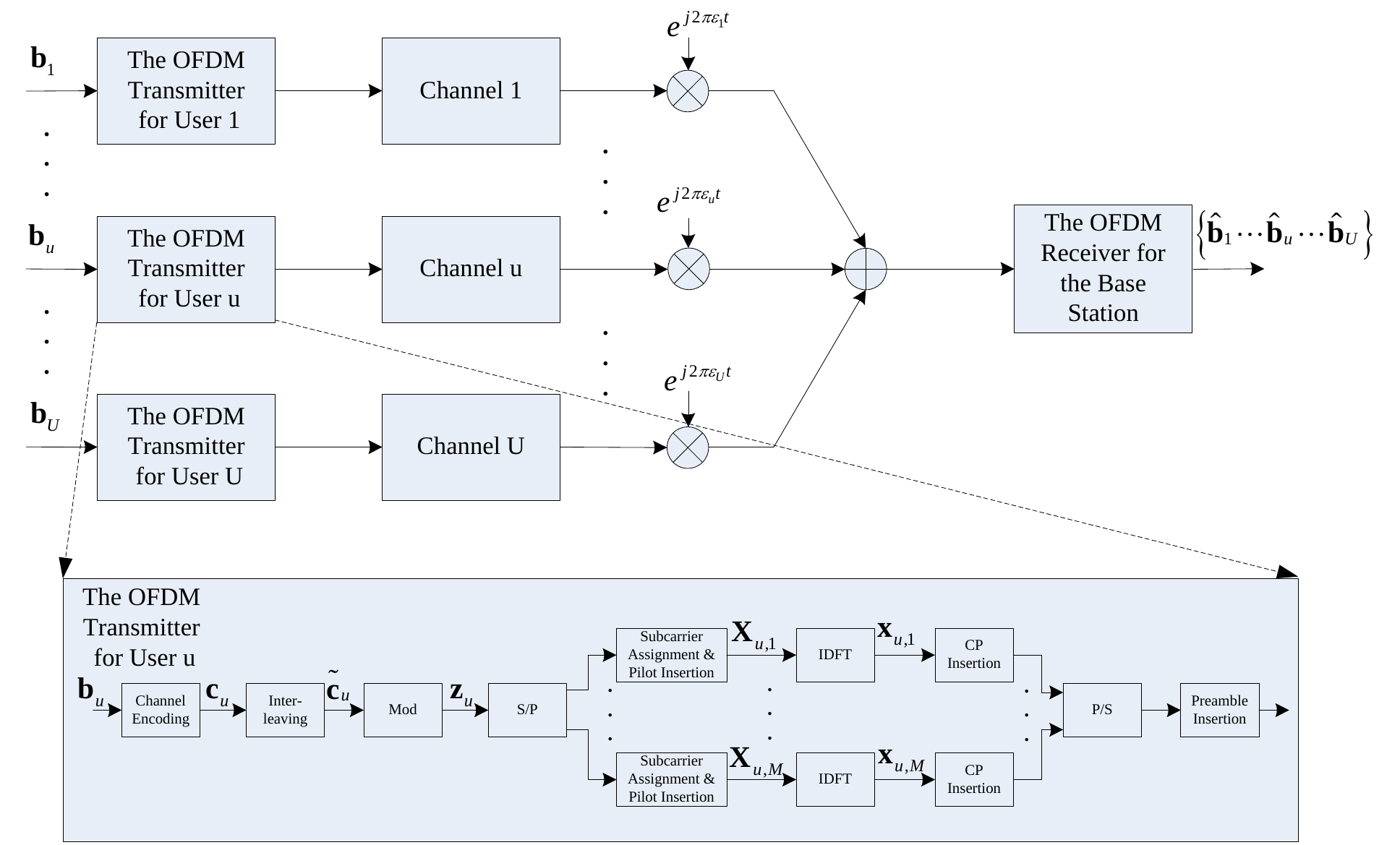}
	\caption{Block diagram of OFDM-IDMA.} \label{block_system_model}
\end{figure*}

Fig.~\ref{block_system_model} shows the block diagram of the system. User $u$, $u \in \left\{ {1,2, \cdots ,U} \right\}$, generates a sequence of $J$ information bits ${{\bf{b}}_u} = {\left[ {{b_{u,1}}{b_{u,2}} \cdots {b_{u,J}}} \right]^T}$. The information bits are then channel-coded into  ${J \mathord{\left/
		{\vphantom {J R}} \right.
		\kern-\nulldelimiterspace} R}$
code bits ${{\bf{c}}_u} = {\left[ {{c_{u,1}}{c_{u,2}} \cdots {c_{u,{J \mathord{\left/
						{\vphantom {J R}} \right.
						\kern-\nulldelimiterspace} R}}}} \right]^T}$, where $R$ is the code rate. We assume all users adopt channel codes that is amenable to decoding by the message passing algorithm \cite{wiberg1996codes, kschischang2001factor}. After channel coding, a user-specific interleaver permutes the sequence of code bits ${{\bf{c}}_u}$  into an interleaved sequence of code bits ${\widetilde{\bf{c}}_u} = {\left[ {{{\widetilde c }_{u,1}}{{\widetilde c}_{u,2}} \cdots {{\widetilde c}_{u,{J \mathord{\left/
						{\vphantom {J R}} \right.
						\kern-\nulldelimiterspace} R}}}} \right]^T}$. The user-specific interleavers together with the channel encoder serve as the \emph{signatures} of the users. Then, ${\widetilde{\bf{c}}_u}$ is modulated to a sequence of complex data symbols ${{\bf{z}}_u} = {\left[ {{z_{u,1}}{z_{u,2}} \cdots {z_{u,{J \mathord{\left/
						{\vphantom {J {RB}}} \right.
						\kern-\nulldelimiterspace} {RB}}}}} \right]^T}$, where $B$ is the number of code bits per complex data symbol. For simplicity, we focus on BPSK modulation in this work. The extension to higher order modulation under the framework of bit-interleaved coded modulation (BICM) \cite{caire1998bit} is straightforward. 

The complex data symbols are transmitted by means of OFDM signaling. OFDM transmits signal on a block-by-block basis. Each OFDM block contains $N$ subcarriers. User $u$ maps each element of ${{\bf{z}}_u}$ to a subcarrier. We denote the $m^{th}$ OFDM block of user $u$  by a length-$N$ vector ${{\bf{X}}_{u,m}} = {\left[ {{X_{u,m,1}}{X_{u,m,2}} \cdots {X_{u,m,N}}} \right]^T}$, where the $i^{th}$ element ${X_{u,m,i}}$, $i \in \left\{ {1,2, \cdots ,N} \right\}$, is the symbol transmitted over the  $i^{th}$ subcarrier in the $m^{th}$ OFDM block of user $u$. 

For channel-parameter estimation, preambles are added to the beginning of the frame, and pilots are carried on selected subcarriers of every OFDM block. In our system, the first $2U$ OFDM blocks at the beginning of the frame are preamble blocks. User $u$ transmits two identical training blocks over preamble block $2u-1$ and $2u$, and nulls its transmission over the other $2U-2$ preamble blocks. Thus, the preambles of different users are mutually orthogonal in the time domain. We stack the OFDM blocks of the overall frame of user $u$ into a length-$MN$ vector ${{\bf{X}}_u} = {\left[ {{\bf{X}}_{u,1}^T{\bf{X}}_{u,2}^T \cdots {\bf{X}}_{u,M}^T} \right]^T}$, where $M$ is the number of OFDM blocks contained in one frame --- the first $2U$ blocks are preambles and the next $M-2U$ blocks are data payload.

We employ the following arrangement for the data and pilot subcarriers. The data subcarriers allocated to different users overlap completely.  The index set of the subcarriers allocated to user data is denoted by $I_D$. The pilot subcarriers allocated to different users do not overlap. The index set of the subcarriers allocated to user $u$'s pilots is denoted by $I_{P,u}$ with  ${I_D} \cap {I_{P,u}} = \emptyset$ for $\forall u$ and ${I_{P,u}} \cap {I_{P,v}} = \emptyset $ for $u \ne v$. User $u$  maps its complex data symbols ${{\bf{z}}_u}$ to data subcarriers according to: ${X_{u,m,i}} = {z_{u,l\left( {m,i} \right)}}$ for $i \in {I_D}$, $2U + 1 \le m \le M$ and $2U + 1 \le m \le M$, where $l\left( {m,i} \right)$ is the index of the complex data symbol assigned to the $i^{th}$ subcarrier of the $m^{th}$ OFDM block. User   $u$ fixes the “known” BPSK symbols on its pilot subcarriers to 1: ${X_{u,m,i}} = 1$ for $i \in {I_{P,u}}$, $2U + 1 \le m \le M$ and $2U + 1 \le m \le M$. On the other users' subcarriers and the guard-band subcarriers, user $u$ transmits dummy null symbols:${X_{u,m,i}} = 0$ for $i \notin {I_D} \cup {I_{P,u}}$, $2U + 1 \le m \le M$ and $2U + 1 \le m \le M$.

We denote the overall function that includes channel encoding, interleaving, modulation, subcarrier mapping, pilot insertion and preamble addition for user $u$  by $C_u$, and we write ${{\bf{X}}_u} = {C_u}\left( {{{\bf{b}}_u}} \right)$ to express that the OFDM symbols in ${\bf{X}}_u$ are mapped from the information bits ${\bf{b}}_u$. 

The OFDM modulation is implemented by an $N$ point IDFT ${{\bf{x}}_{u,m}} = {{\bf{F}}^H}{{\bf{X}}_{u,m}}$, where ${{\bf{x}}_{u,m}} = {\left[ {{x_{u,m,1}}{x_{u,m,2}} \cdots {x_{u,m,N}}} \right]^T}$ is the vector of time-domain samples, and $\bf{F}$ is the $N \times N$ DFT matrix whose ${\left( {p,q} \right)^{th}}$ entry is given by ${{{{\mathop{\rm e}\nolimits} ^{ - j2\pi {{\left( {p - 1} \right)\left( {q - \left( {{{1 + N} \mathord{\left/
										{\vphantom {{1 + N} 2}} \right.
										\kern-\nulldelimiterspace} 2}} \right)} \right)} \mathord{\left/
						{\vphantom {{\left( {p - 1} \right)\left( {q - \left( {{{1 + N} \mathord{\left/
													{\vphantom {{1 + N} 2}} \right.
													\kern-\nulldelimiterspace} 2}} \right)} \right)} N}} \right.
						\kern-\nulldelimiterspace} N}}}} \mathord{\left/
		{\vphantom {{{{\mathop{\rm e}\nolimits} ^{ - j2\pi {{\left( {p - 1} \right)\left( {q - \left( {{{1 + N} \mathord{\left/
													{\vphantom {{1 + N} 2}} \right.
													\kern-\nulldelimiterspace} 2}} \right)} \right)} \mathord{\left/
									{\vphantom {{\left( {p - 1} \right)\left( {q - \left( {{{1 + N} \mathord{\left/
																{\vphantom {{1 + N} 2}} \right.
																\kern-\nulldelimiterspace} 2}} \right)} \right)} N}} \right.
									\kern-\nulldelimiterspace} N}}}} {\sqrt N }}} \right.
		\kern-\nulldelimiterspace} {\sqrt N }}$, $1 \le p,q \le N$, $1 \le p,q \le N$.
We stack all the time-domain sample vectors of the whole frame into a length-$MN$ vector ${{\bf{x}}_u} = {\left[ {{\bf{x}}_{u,1}^T{\bf{x}}_{u,2}^T \cdots {\bf{x}}_{u,M}^T} \right]^T}$.

To overcome the delay spread of multipath channels, which causes inter-block interferences (IBI), the time-domain samples of each OFDM symbol is preceded by a cyclic prefix (CP). We denote the length of the CP by ${N_{cp}}$. Therefore, each OFDM block includes ${N_s} = {N_{cp}} + N$ time-domain samples $\left\{ {{x_{u,m,i}}} \right\}_{i =  - {N_{cp}} + 1}^N$, where ${x_{u,m,i}} = {x_{u,m,i + N}}$ for $i \in \left\{ { - {N_{cp}} + 1, \cdots ,0} \right\}$. The time-domain samples are converted to signal waveform via a digital-to-analog converter (DAC). The $U$ users simultaneously transmit their signal waveforms on their respective multipath channels. The base station receives their overlapped signal waveforms.

\subsection{Channel Model} 

We denote the overall discrete time-domain channel impulse response of user $u$ that captures the effects of both the physical channel and the transmit/receive filters by ${\widetilde{\bf{h}}_u} = {\left[ {{h_{u,1}}{h_{u,2}} \cdots {h_{u,{L_u}}}} \right]^T}$, where ${h_{u,l}}$ is the $l^{th}$ discrete tap of the multipath channel of user $u$, $l \in \left\{ {1, \cdots ,{L_u}} \right\}$, and $L_u$ is the maximum channel delay spread. We assume that the channels remain static over the transmission time of one frame. This assumption is valid in scenarios where users exhibit low mobility \cite{liu2004parameter}. We model the channel taps as mutually independent complex Gaussian variables with zero mean and an exponentially decaying power delay profile: ${\rm{E}}\left\{ {{{\left| {{h_{u,l}}} \right|}^2}} \right\} = \beta \exp \left( { - {{\left( {l - 1} \right)} \mathord{\left/
			{\vphantom {{\left( {l - 1} \right)} {{L_u}}}} \right.
			\kern-\nulldelimiterspace} {{L_u}}}} \right)$, where $\beta $ is a normalization factor to ensure that the average channel energy is one.

\subsection{Receive Signal}

The duration of a time-domain sample is denoted by ${T_s}$. The the timing mismatch between user $u$ and the base station  is denoted by ${\tau _u}$. For convenience, we define the \emph{zero reference time} as the first-channel-path arrival time of user 1 (${\tau _1} = 0$), and assume without loss of generality that the first-channel-path arrival times of all other users are later than that of user 1 (${\tau _u} > 0$, $u \in \left\{ {2, \cdots ,U} \right\}$). Following \cite{morelli2004timing, pun2007iterative}, we decompose ${\tau _u}$  into an integer part plus a fractional part with respect to the sample duration $T_s$: ${\tau _u} = \left( {{\mu _u} + {\delta _u}} \right){T_s}$, where ${\mu _u} = \left\lfloor {{{{\tau _u}} \mathord{\left/
			{\vphantom {{{\tau _u}} {{T_s}}}} \right.
			\kern-\nulldelimiterspace} {{T_s}}}} \right\rfloor $ is the integer part and    ${\delta _u} = {{{\tau _u}} \mathord{\left/
		{\vphantom {{{\tau _u}} {{T_s}}}} \right.
		\kern-\nulldelimiterspace} {{T_s}}} - {\mu _u}$
is the fractional part. The fractional part can be incorporated into the channel impulse response, thus we will not consider it going forward. We can avoid IBI in the system by assuming that the system satisfies a \emph{loose time synchronization requirement}, specified as ${\max _u}\left\{ {{\mu _u} + {L_u}} \right\} \le {N_{cp}}$. As long as this requirement is satisfied, symbol misalignment in the time domain does not affect the subcarrier-by-subcarrier channel decoding in the frequency domain other than introducing relative phase offsets between the signals of the users --- in particular, there is no symbol misalignment in the frequency domain and signals on different subcarriers are isolated from one another.  Note that there is no need to explicitly estimate the timing parameters $\left\{ {{\tau _u}} \right\}_{u = 2}^U$. To meet the loose time synchronization requirement, the base station can broadcast a downlink beacon to prompt the users to transmit together \cite{morelli2004timing, pun2007iterative, lu2012implementation, lu2013real}. 

The presence of CFOs destroys the perfect orthogonalities among subcarriers. A main focus of this paper is the study of the effect of CFOs (i.e., the frequency asynchrony problem). After analog-to-digital conversion and removal of the first $N_{cp}$ CP samples (counting from the first sample of the first channel path of user 1) of every OFDM block, the received discrete time-domain samples at the base station can be expressed as
\begin{equation}\label{sm1}
	{{\bf{r}}_m} = \sum\limits_{u = 1}^U {{e^{j{\theta _{u,m}}}}{\bf{\Gamma }}\left( {{\varepsilon _u}} \right){{\bf{F}}^H}{\bf{D}}\left( {{{\bf{X}}_{u,m}}} \right){\bf{F}}{{\bf{h}}_u}}  + {{\bf{n}}_m}
\end{equation}
for $m = 1,2, \cdots ,M$, where

\begin{itemize}
	\item  	${{\bf{r}}_m} = {\left[ {{r_{m,1}}{r_{m,2}} \cdots {r_{m,N}}} \right]^T}$ is the vector of the $N$ discrete samples of the $m^{th}$  received OFDM block; ${{\bf{n}}_m} = {\left[ {{n_{m,1}}{n_{m,2}} \cdots {n_{m,N}}} \right]^T}$ is the vector of the complex white Gaussian noises with zero-mean and variance of $\sigma _n^2$;

	\item ${\varepsilon _u}$ is the CFO (normalized to subcarrier space) between the base station and user $u$; ${\bf{\Gamma }}\left( {{\varepsilon _u}} \right) = diag\left\{ {\left[ {1,{e^{j{{2\pi {\varepsilon _u}} \mathord{\left/
							{\vphantom {{2\pi {\varepsilon _u}} N}} \right.
							\kern-\nulldelimiterspace} N}}}, \cdots ,{e^{j{{2\pi {\varepsilon _u}\left( {N - 1} \right)} \mathord{\left/
							{\vphantom {{2\pi {\varepsilon _u}\left( {N - 1} \right)} N}} \right.
							\kern-\nulldelimiterspace} N}}}} \right]} \right\}$ is the diagonal matrix that captures the time-domain effect of CFO on the $m^{th}$ block of user $u$;

	\item ${\theta _{u,m}} = 2\pi {\varepsilon _u}{{\left( {{N_{cp}} + m{N_s}} \right)} \mathord{\left/
			{\vphantom {{\left( {{N_{cp}} + m{N_s}} \right)} N}} \right.
			\kern-\nulldelimiterspace} N}$ is the accumulated phase drift caused by CFO ${\varepsilon _u}$ on the $m^{th}$ block of user $u$; 
	
	\item ${\bf{D}}\left( {{{\bf{X}}_{u,m}}} \right) = diag\left( {{{\bf{X}}_{u,m}}} \right)$ is the diagonal matrix with transmitted frequency-domain symbols ${{\bf{X}}_{u,m}}$ as its diagonal elements;
	
	\item	${{\bf{h}}_u} \buildrel \Delta \over = {\left[ {{\bf{0}}_{{\mu _u}}^T,\widetilde{\bf{h}}_u^T,{\bf{0}}_{N - {L_u} - {\mu _u}}^T} \right]^T}$ is the length-$N$ vector that captures both the discrete CIR and the time asynchrony of node $u$.
	
\end{itemize}

For OFDM systems, channel decoding is performed in the frequency domain. The frequency-domain sample vector of the $m^{th}$ OFDM symbol, ${{\bf{R}}_m} = {\left[ {{R_{m,1}}{R_{m,2}} \cdots {R_{m,N}}} \right]^T} = {\bf{F}}{{\bf{r}}_m}$, is given by  
\begin{equation} \label{sm2}
	\begin{array}{c}
		{{\bf{R}}_m}{\rm{ }} = \sum\limits_{u = 1}^U {{e^{j{\theta _{u,m}}}}\underbrace {{\bf{F\Gamma }}\left( {{\varepsilon _u}} \right){{\bf{F}}^H}}_{ \buildrel \Delta \over = {\bf{\Xi }}\left( {{\varepsilon _u}} \right)}{\bf{D}}\left( {{{\bf{X}}_{u,m}}} \right)\underbrace {{\bf{F}}{{\bf{h}}_u}}_{ \buildrel \Delta \over = {{\bf{H}}_u}}}  + \underbrace {{\bf{F}}{{\bf{n}}_m}}_{ \buildrel \Delta \over = {{\bf{N}}_m}} \\ 
		= \sum\limits_{u = 1}^U {{e^{j{\theta _{u,m}}}}{\bf{\Xi }}\left( {{\varepsilon _u}} \right){\bf{D}}\left( {{{\bf{X}}_{u,m}}} \right){{\bf{H}}_u}}  + {{\bf{N}}_m} \\ 
	\end{array}
\end{equation}                       
where ${{\bf{H}}_u} \buildrel \Delta \over = {\bf{F}}{{\bf{h}}_u}$ is the vector of the frequency channel responses for user $u$, ${{\bf{N}}_m} \buildrel \Delta \over = {\bf{F}}{{\bf{n}}_m}$ is the vector of frequency-domain noises, ${\bf{\Xi }}\left( {{\varepsilon _u}} \right) \buildrel \Delta \over = {\bf{F\Gamma }}\left( {{\varepsilon _u}} \right){{\bf{F}}^H}$ is the matrix that transforms the effect of CFO ${\varepsilon _u}$ in the time domain to ICI in the frequency domain. It can be shown that ${\bf{\Xi }}\left( {{\varepsilon _u} = 0} \right) = {\bf{I}}$; ${\bf{\Xi }}\left( {{\varepsilon _u} \ne 0} \right) = \lambda {\bf{I}} + {\bf{\Pi }}\left( {{\varepsilon _u} \ne 0} \right)$, where $\lambda $ is a common attenuation factor across all subcarriers, and ${\bf{\Pi }}\left( {{\varepsilon _u} \ne 0} \right)$ is the ICI component \cite{rugini2005ber}.

We stack the $M$ length-$N$ time-domain (frequency-domain) sample vectors into an overall length-$MN$ sample vector ${\bf{r}} \buildrel \Delta \over = {\left[ {{\bf{r}}_1^T{\bf{r}}_2^T \cdots {\bf{r}}_M^T} \right]^T}$ (${\bf{R}} \buildrel \Delta \over = {\left[ {{\bf{R}}_1^T{\bf{R}}_2^T \cdots {\bf{R}}_M^T} \right]^T}$). The receiver then undertakes two tasks: (i) estimation of channel parameters $\left\{ {{\varepsilon _u},{{\bf{h}}_u}} \right\}_{u = 1}^U$; (ii)  decoding of multiuser information bit sequences  $\left\{ {{{\bf{b}}_u}} \right\}_{u = 1}^U$
.  

With reference to the signal model in (\ref{sm2}), we can observe that CFO causes two negative effects on the frequency-domain signal: (i) the drifting of the signal phase over time; (ii) the inter-carrier interferences (ICI) among different subcarriers.

For conventional point-to-point OFDM systems, the receiver can first estimate the CFO and the channel gain from the preamble and then compensate for the CFO for the whole frame in the time domain. That is, it attempts to remove the CFO in the signal using the estimated CFO. After that, it performs standard channel decoding in the frequency domain. Estimation error may leave behind an uncompensated residual CFO. If the estimation is accurate enough, the residual CFO will be small, and the remaining CFO-induced ICI will also be small. We can treat the residual ICI as additional noise that lowers the effective signal-to-noise ratio (SNR) slightly \cite{speth1999optimum}, but standard channel encoding/decoding can still be employed to ensure communication reliability \cite{sathananathan2001forward}. This CFO compensation alone, however, cannot overcome the drifting phase. Even a small residual CFO can lead to large phase drifts accumulated over time. Standard channel decoding will fail if we ignore the phase drifts. Therefore, point-to-point OFDM systems typically employ pilot subcarriers in the OFDM block to track signal phase so that it can be corrected.  \cite{liu2004parameter}.

The above design principle, however, does not work for the multiuser OFDM-IDMA system. In particular, there are multiple CFOs in OFDM-IDMA, one for the signal of each user. First of all, accurately estimating these CFOs and the channel gains is demanding. More fundamentally, compensating for these CFOs simultaneously in one shot is impossible even if the CFOs could be estimated without errors (by contrast, for conventional point-to-point systems, total CFO removal is possible given perfect CFO estimation). This is because removing one of the CFOs in the received signal will necessarily leave behind some remaining CFOs for the other CFOs, unless the CFOs of different users were exactly the same to begin with. As a consequence, the CFO-induced ICI inevitably remains in the frequency domain. Instead of channel-parameter estimation, followed by CFO compensation, followed by channel decoding, that are typical in many single-user systems, an iterative method is called for in a multiuser system such as OFDM-IDMA. In what follows, we propose a SAGE-and-ECM framework to address the joint problem of multiuser channel-parameter estimation, CFO compensation and channel decoding.

\section{Multiuser Joint Channel-Parameter Estimation, CFO Compensation and Channel Decoding}

To improve the overall performance of the OFDM-IDMA system, we make use of the preambles, the pilots as well as the data payload of the received signal to jointly estimate channel parameters and decode information bits.  

For OFDMA receivers, \cite{pun2007iterative} developed an iterative solution that uses SAGE and ECM.  However, \cite{pun2007iterative} did not consider the impact of channel coding, the good performance of which necessitates a total recast of the algorithmic framework. In this work, we explore the proper algorithmic framework of SAGE and ECM for channel-coded OFDM-IDMA systems. We argue that incorporating channel decoding by simply extending the framework of \cite{pun2007iterative} leads to subpar performance. 

We formulate our target problem as a multidimensional estimation problem in Section III.A. Section III.B gives an overview of a SAGE-based signal decomposition method. As in \cite{pun2007iterative}, we apply SAGE and break down the overall problem into $U$ sub-problems, one for each user, by decomposing the received signal into $U$ signal components. Readers who are familiar with SAGE decomposition can quickly go through Section III.B to familiarize themselves with our notations. Section III.C zooms in to the ECM-based joint channel-parameter estimation, CFO compensation and channel decoding problem within the $U$ sub-problems. The main contribution of our work is contained in Section III.C. In particular, a key issue is how to incorporate message passing algorithm for channel decoding into the joint framework. In that regard, two message passing strategies are possible for channel decoding: sum-product and min-sum message passing \cite{wiberg1996codes, kschischang2001factor}. In our framework, the ECM algorithm gives the sum-product channel decoding. We show that the integration of sum-product channel decoding with channel-parameter estimation leads to better system performance than min-sum channel decoding. However, for compatibility with the channel-parameter estimation part, the sum-product channel decoding cannot be applied directly: a transformation of the soft information from the time domain to the frequency domain is needed. We develop a new technique called “soft IDFT” to realize an overall compatible algorithmic framework. Exact computation of soft IDFT, however, is complex. Here, we obtain an effective approximate solution by Gaussian message passing.

\subsection{Problem Statement}

As expressed by the signal model in (\ref{sm1}), the only unknown channel parameters are the CFOs $\left\{ {{\varepsilon _u}} \right\}_{u = 1}^U$ and the channel gains $\left\{ {{{\bf{h}}_u}} \right\}_{u = 1}^U$. In (\ref{sm1}), the  phase drift ${\theta _{u,m}} = 2\pi {\varepsilon _u}{{\left( {{N_{cp}} + m{N_s}} \right)} \mathord{\left/
		{\vphantom {{\left( {{N_{cp}} + m{N_s}} \right)} N}} \right.
		\kern-\nulldelimiterspace} N}$ is not independent and is a function of CFO ${\varepsilon _u}$. However, if we only estimate ${\varepsilon _u}$ and derive the estimated phase drift ${\theta _{u,m}}$ based on the estimated ${\varepsilon _u}$, even a tiny estimation error in ${\varepsilon _u}$ will accumulate to large estimation errors in ${\theta _{u,m}}$ for later blocks $m$. Furthermore, the expression   ${\theta _{u,m}} = 2\pi {\varepsilon _u}{{\left( {{N_{cp}} + m{N_s}} \right)} \mathord{\left/
		{\vphantom {{\left( {{N_{cp}} + m{N_s}} \right)} N}} \right.
		\kern-\nulldelimiterspace} N}$ assumes that the signals of different users do not incur different phase noises and the phase ${\theta _{u,m}}$ is strictly due to that of the CFO. In practice, the phase drift is not due to CFO alone but also due to phase noise that behaves like a random walk \cite{demir2000phase}. Thus, in this paper, for a robust system, we do not make use of the expression ${\theta _{u,m}} = 2\pi {\varepsilon _u}{{\left( {{N_{cp}} + m{N_s}} \right)} \mathord{\left/
		{\vphantom {{\left( {{N_{cp}} + m{N_s}} \right)} N}} \right.
		\kern-\nulldelimiterspace} N}$ when estimating ${\theta _{u,m}}$: we assume that $\left\{ {{\theta _{u,m}}} \right\}$ are independent for different OFDM blocks and different users, and estimate ${\theta _{u,m}}$ of each block independently. The advantage of this scheme is that phase errors due to estimated CFO errors are not cumulative over  blocks and random phase noise can be taken into account. 

With the above, the overall unknown variables in the system are $\left\{ {{\varepsilon _u},\left\{ {{\theta _{u,m}}} \right\}_{m = 1}^M,{{\bf{h}}_u},{{\bf{X}}_u}} \right\}_{u = 1}^U$.

According to the maximum likelihood (ML) principle, we can express the objective of the joint channel-parameter estimation and channel decoding problem as
\begin{equation}\label{problem}
\begin{array}{l}
\left( {\left\{ {{{\widehat\varepsilon }_u},\left\{ {{{\widehat\theta }_{u,m}}} \right\}_{m = 1}^M,{{\widehat{\bf{h}}}_u},{{\widehat{\bf{X}}}_u}} \right\}_{u = 1}^U} \right) \\ 
=   \arg \mathop {\max }\limits_{\left\{ {{\varepsilon _u},\left\{ {{\theta _{u,m}}} \right\}_{m = 1}^M,{{\bf{h}}_u},{\bf{X}} \in {C_u}} \right\}_{u = 1}^U} \left\{ \begin{array}{l}
\\ 
\\ 
\end{array} \right. \\ 
\;\;\;\;\;\;\;\;\;\;\;\;\;\;\;\;\;\;\;\;\;\;\; \left. {\log p\left( {{\bf{r}}\left| {\left\{ {{\varepsilon _u},\left\{ {{\theta _{u,m}}} \right\}_{m = 1}^M,{{\bf{h}}_u},{{\bf{X}}_u}} \right\}_{u = 1}^U} \right.} \right)} \right\} \\ 
= \arg \mathop {\min }\limits_{\left\{ {{\varepsilon _u},\left\{ {{\theta _{u,m}}} \right\}_{m = 1}^M,{{\bf{h}}_u},{\bf{X}} \in {C_u}} \right\}_{u = 1}^U} \left\{ \begin{array}{l}
\\ 
\\ 
\end{array} \right. \\ 
\;\;\;\;\;\;\;\;\;\;\;\; \left. {\sum\limits_{m = 1}^M {{{\left\| {{{\bf{r}}_m} - \sum\limits_{u = 1}^U {{e^{j{\theta _{u,m}}}}{\bf{\Gamma }}\left( {{\varepsilon _u}} \right){{\bf{F}}^H}{\bf{D}}\left( {{{\bf{X}}_{u,m}}} \right){\bf{F}}{{\bf{h}}_u}} } \right\|}^2}} } \right\} \\ 
\end{array}
\end{equation}
With regard to (\ref{problem}), we emphasize that since the sequence of transmit symbols is generated from the sequence of the original information bits via a one-to-one mapping, i.e., $\left\{ {{{\bf{X}}_u} = {C_u}\left( {{{\bf{b}}_u}} \right)} \right\}_{u = 1}^U$, decoding the transmit symbols $\left\{ {{{\bf{X}}_u}} \right\}_{u = 1}^U$ is equivalent to decoding the information bits $\left\{ {{{\bf{b}}_u}} \right\}_{u = 1}^U$. Directly solving the ML problem (\ref{problem}) is intractable because the exhaustive search over the multi-dimensional space of  $\left\{ {{\varepsilon _u},\left\{ {{\theta _{u,m}}} \right\}_{m = 1}^M,{{\bf{h}}_u},{{\bf{X}}_u}} \right\}_{u = 1}^U$ is prohibitively complex.

\subsection{Preliminary for the Signal Decomposition Using SAGE}
In \cite{feder1988parameter}, the authors solved the problem of multiple parameter estimation (but not channel decoding) using a iterative method. Later on, the method of \cite{feder1988parameter} evolved into the SAGE algorithm \cite{sage1994}. The key idea is to decompose the received overlapping signal into several signal components. Subsequently, \cite{pun2007iterative} applied the method of \cite{feder1988parameter} to OFDMA systems.  Since the signal decomposition is performed in the time domain, where the user signals overlap completely (for both OFDM-IDMA and OFDMA), we can directly apply this SAGE-based signal decomposition to our OFDM-IDMA system. This subsection extends the SAGE signal decomposition in \cite{feder1988parameter} to incorporate channel decoding.

First, the overall estimation problem (\ref{problem}) is decomposed into $U$ sub-problems \cite{feder1988parameter, pun2007iterative}.  For the $m^{th}$ OFDM block, the signal component of user $u$ is defined to be 
\begin{equation}
	{{\bf{r}}_{u,m}} \buildrel \Delta \over = {e^{j{\theta _{u,m}}}}{\bf{\Gamma }}\left( {{\varepsilon _u}} \right){{\bf{F}}^H}{\bf{D}}\left( {{{\bf{X}}_{u,m}}} \right){\bf{F}}{{\bf{h}}_u} + {{\bf{n}}_{u,m}}
\end{equation}
where $\left\{ {{{\bf{n}}_{u,m}}} \right\}_{u = 1}^U$ are obtained by arbitrarily decomposing the total noise vector ${{\bf{n}}_m}$ into $U$ circularly symmetric and statistically independent noise component vectors that satisfy ${{\bf{n}}_m} = \sum\nolimits_{u = 1}^U {{{\bf{n}}_{u,m}}}$  \cite{feder1988parameter}.  The received signal can then be written as  
\begin{equation}\label{decomp}
	{{\bf{r}}_m} = \sum\nolimits_{u = 1}^U {{{\bf{r}}_{u,m}}} 
\end{equation}
where ${{\bf{r}}_m}$ is decomposed into $U$  components $\left\{ {{{\bf{r}}_{u,m}}} \right\}_{u = 1}^U$, with  each being exclusively related to one user. As with the stacking of the overall received signals ${\bf{r}} = {\left[ {{\bf{r}}_1^T{\bf{r}}_2^T \cdots {\bf{r}}_M^T} \right]^T}$ in the whole frame, we can also stack the $M$ signal component vectors of user $u$ into an overall vector ${{\bf{r}}_u} \buildrel \Delta \over = {\left[ {{\bf{r}}_{u,1}^T{\bf{r}}_{u,2}^T \cdots {\bf{r}}_{u,M}^T} \right]^T}$, $u \in \left\{ {1,2, \cdots ,U} \right\}$. In the terminology of SAGE, $\bf{r}$ is the observed data and  $\left\{ {{{\bf{r}}_u}} \right\}_{u = 1}^U$ is the complete data.

SAGE tries to find the ML estimates for $\left\{ {{\varepsilon _u},} \right.$ $\left. {\left\{ {{\theta _{u,m}}} \right\}_{m = 1}^M,{{\bf{h}}_u},{{\bf{X}}_u}} \right\}_{u = 1}^U$ iteratively.\footnote{This means that the setup of SAGE here treats all variables $\left\{ {{\varepsilon _u},} \right.$ $\left. {\left\{ {{\theta _{u,m}}} \right\}_{m = 1}^M,{{\bf{h}}_u},{{\bf{X}}_u}} \right\}_{u = 1}^U$, including $\left\{ {{{\bf{X}}_u}} \right\}_{u = 1}^U$ as the parameters; there is no hidden data in the setup \cite{sage1994, feder1988parameter}.} Let $\left\{ {\widehat\varepsilon _u^{\left( k \right)},\left\{ {\widehat\theta _{u,m}^{\left( k \right)}} \right\}_{m = 1}^M,\widehat{\bf{h}}_u^{\left( k \right)},\widehat{\bf{X}}_u^{\left( k \right)}} \right\}_{u = 1}^U$ be the updated estimates after the $k^{th}$ SAGE iteration, where $k = 1,2, \cdots K$
and $\left\{ {\widehat\varepsilon _u^{\left( 0 \right)},\left\{ {\widehat\theta _{u,m}^{\left( 0 \right)}} \right\}_{m = 1}^M,\widehat{\bf{h}}_u^{\left( 0 \right)},\widehat{\bf{X}}_u^{\left( 0 \right)}} \right\}_{u = 1}^U$ are the initial estimates. The initial estimates for the CFOs and the channel gains are obtained from the orthogonal preambles; the initial estimates for the phase drifts are set to zeros; the initial estimates for the transmit symbols are set to $\pm 1$ randomly. Each SAGE iteration consists of $U$ cycles; the variables of user $u$ are updated in the $u^{th}$ cycle given that the variables of all other users are fixed to their last estimates.

With the initial estimates $\left\{ {\widehat\varepsilon _u^{\left( 0 \right)},\left\{ {\widehat\theta _{u,m}^{\left( 0 \right)}} \right\}_{m = 1}^M,\widehat{\bf{h}}_u^{\left( 0 \right)},\widehat{\bf{X}}_u^{\left( 0 \right)}} \right\}_{u = 1}^U$,  the SAGE algorithm first computes the initial estimates for the individual user signal components:
\begin{equation}
	\widehat{\bf{r}}_{u,m}^{\left( 0 \right)} = {e^{j\widehat\theta _{u,m}^{\left( 0 \right)}}}{\bf{\Gamma }}\left( {\widehat\varepsilon _u^{\left( 0 \right)}} \right){{\bf{F}}^H}{\bf{D}}\left( {\widehat{\bf{X}}_u^{\left( 0 \right)}} \right){\bf{F}}\widehat{\bf{h}}_u^{\left( 0 \right)}
\end{equation}
where $m = 1,2, \cdots ,M$ and $u \in \left\{ {1,2, \cdots ,U} \right\}$. The $u^{th}$ cycle of the $k^{th}$ SAGE iteration includes an E-step and an M-step \cite{feder1988parameter} as follows:
\\

\noindent \textbf{E-step of SAGE}:
\\
Compute the tentative estimate for the signal component of user $u$:
\begin{equation}\label{e-sage}
	\widehat{\bf{r}}_{u,m}^{\left( k \right)} = {{\bf{r}}_m} - \sum\limits_{v = 1}^{u - 1} {\widehat{\bf{r}}_{v,m}^{\left( k \right)}}  - \sum\limits_{v = u + 1}^U {\widehat{\bf{r}}_{v,m}^{\left( {k - 1} \right)}}
\end{equation}
where $m = 1,2, \cdots ,M$, and $\sum\nolimits_v^u { = 0}$ if $u < v$. Note that this computation is based on the signal decomposition in (\ref{decomp}). We stack the estimates for all the OFDM blocks into an overall vector $\widehat{\bf{r}}_u^{\left( k \right)} \buildrel \Delta \over = {\left[ {\widehat{\bf{r}}_{u,1}^{\left( k \right)T}\widehat{\bf{r}}_{u,2}^{\left( k \right)T} \cdots \widehat{\bf{r}}_{u,M}^{\left( k \right)T}} \right]^T}$. 
\\

\noindent \textbf{M-step of SAGE}:
\\
Update the estimates for the variables of user $u$:
\begin{equation}\label{subprob}
\begin{array}{l}
\left( {\widehat\varepsilon _u^{\left( k \right)},\left\{ {\widehat\theta _{u,m}^{\left( k \right)}} \right\}_{m = 1}^M,\widehat{\bf{h}}_u^{\left( k \right)},\widehat{\bf{X}}_u^{\left( k \right)}} \right) \\ 
= \arg \mathop {\max }\limits_{\left( {{\varepsilon _u},\left\{ {{\theta _{u,m}}} \right\}_{m = 1}^M,{{\bf{h}}_u},{{\bf{X}}_u} \in {C_u}} \right)} \left\{ \begin{array}{l}
\\ 
\\ 
\end{array} \right. \\ 
\;\;\;\;\;\;\;\;\;\;\;\;\;\;\;\;\;\;\;\;\;\;\;\;\;\;\;\;\;\;  \left. {\log p\left( {\widehat{\bf{r}}_u^{\left( k \right)}\left| {{\varepsilon _u},\left\{ {{\theta _{u,m}}} \right\}_{m = 1}^M,{{\bf{h}}_u},{{\bf{X}}_u}} \right.} \right)} \right\} \\ 
= \arg \mathop {\min }\limits_{\left( {{\varepsilon _u},\left\{ {{\theta _{u,m}}} \right\}_{m = 1}^M,{{\bf{h}}_u},{{\bf{X}}_u} \in {C_u}} \right)} \left\{ \begin{array}{l}
\\ 
\\ 
\end{array} \right. \\ 
\;\;\;\;\;\;\;\;\;\;\;\;\;\;\; \left. {\sum\limits_{m = 1}^M {{{\left\| {\widehat{\bf{r}}_{m,u}^{\left( k \right)} - {e^{j{\theta _{u,m}}}}{\bf{\Gamma }}\left( {{\varepsilon _u}} \right){{\bf{F}}^H}{\bf{D}}\left( {{{\bf{X}}_{u,m}}} \right){\bf{F}}{{\bf{h}}_u}} \right\|}^2}} } \right\} \\ 
\end{array}
\end{equation}
After the M-step, we then reconstruct the estimate for the signal component of user $u$ using the updated variable estimates
\begin{equation}
	\widehat{\bf{r}}_{u,m}^{\left( 0 \right)} = {e^{j\widehat\theta _{u,m}^{\left( k \right)}}}{\bf{\Gamma }}\left( {\widehat\varepsilon _u^{\left( k \right)}} \right){{\bf{F}}^H}{\bf{D}}\left( {\widehat{\bf{X}}_u^{\left( k \right)}} \right){\bf{F}}\widehat{\bf{h}}_u^{\left( k \right)}
\end{equation}
for $m = 1,2, \cdots ,M$. This completes the $u^{th}$ cycle of the $k^{th}$  SAGE iteration; we then proceed to the cycle of the next user.

After the variables of all users are updated to $\left\{ {\widehat\varepsilon _u^{\left( k \right)},\left\{ {\widehat\theta _{u,m}^{\left( k \right)}} \right\}_{m = 1}^M,\widehat{\bf{h}}_u^{\left( k \right)}} \right.$ $,\left. {\widehat{\bf{X}}_u^{\left( k \right)}} \right\}_{u = 1}^U$, we then proceed to the next SAGE iteration. When the number of iteration $k$ reaches a preset maximum limit $K$, we terminate the SAGE algorithm after obtaining the final variable estimates $\left\{ {\widehat\varepsilon _u^{\left( K \right)},\left\{ {\widehat\theta _{u,m}^{\left( K \right)}} \right\}_{m = 1}^M,\widehat{\bf{h}}_u^{\left( K \right)},\widehat{\bf{X}}_u^{\left( K \right)}} \right\}_{u = 1}^U$. According to the theory of SAGE \cite{sage1994}, it is expected that the final variable estimates will converge to the global optimal as required by the ML estimation in (\ref{problem}).

As seen above, SAGE decomposes the multiuser problem of joint channel-parameter estimation and channel decoding in  (\ref{problem}) into  $U$ single-user problems in (\ref{subprob}). The complexity of the multiuser problem is reduced substantially. However, the computation involved in the single-user sub-problems as expressed in  (\ref{subprob}) is still  non-trivial. For each user, we need to solve a multi-dimensional problem associated with simultaneously estimating $\left\{ {{\varepsilon _u},\left\{ {{\theta _{u,m}}} \right\}_{m = 1}^M,{{\bf{h}}_u},{{\bf{X}}_u}} \right\}$. This is of high complexity, particularly if the data symbols  $\left\{ {{{\bf{X}}_u}} \right\}_{u = 1}^U$  are channel-coded symbols and we want to exploit the correlations among the symbols induced by channel coding to optimize our estimation.  

A simplified approach to solve  (\ref{subprob}) is to estimate the variables in $\left\{ {{\varepsilon _u},\left\{ {{\theta _{u,m}}} \right\}_{m = 1}^M,{{\bf{h}}_u},{{\bf{X}}_u}} \right\}$   one at a time  sequentially and iteratively. When one variable is under estimation, all other variables are fixed to their estimates from the last iteration. This approach is straightforward with an important caveat: it does not use the information obtained from channel decoding in an optimal way.  We will further elaborate on this simplified approach in Section III.D and will treat it as a benchmark for evaluating our approach to be presented in Section III.C.  To make better use of the information from channel decoding, Section III.C will construct a more comprehensive approach to our problem using the ECM algorithm.

\subsection{Joint Channel-Parameter Estimation and Channel Decoding Using ECM}

ECM is a variant of the expectation maximization (EM) algorithm, a general iterative algorithm for finding the ML estimates of parameters in a statistical model with hidden data \cite{dempster1977EM}. EM updates all parameters in the model simultaneously at each iteration. This requires EM to operate in a multi-dimensional space. To reduce complexity, ECM updates the parameters sequentially. At each stage of the update sequence, ECM updates just one parameter, fixing the other parameters to their last estimates \cite{mclachlan2007algorithm}.

Within the framework of ECM, we could assign the role of hidden data and the role of parameters to the variables  $\left\{ {{\varepsilon _u},\left\{ {{\theta _{u,m}}} \right\}_{m = 1}^M,{{\bf{h}}_u},{{\bf{X}}_u}} \right\}$ in different ways, each yielding a different implementation. In that sense, it is more general than (\ref{subprob}), in which all variables are treated as parameters.

In this paper and the rest of this section, we focus on the assignment that treats ${{\bf{X}}_u}$ as the hidden data, and $\left\{ {{\varepsilon _u},\left\{ {{\theta _{u,m}}} \right\}_{m = 1}^M,{{\bf{h}}_u}} \right\}$ as the parameters.\footnote{The assignment that treats ${{\bf{h}}_u}$ as the hidden data and $\left\{ {{\varepsilon _u},\left\{ {{\theta _{u,m}}} \right\}_{m = 1}^M,{{\bf{X}}_u}} \right\}$ as the parameters is discussed in Appendix C.} With this assignment, we can incorporate sum-product channel decoding into the joint framework. Now, ECM seeks to solve the following maximization problem:
\begin{equation}\label{ecmprob}
\begin{array}{l}
\left( {\widehat\varepsilon _u^{\left( k \right)},\left\{ {\widehat\theta _{u,m}^{\left( k \right)}} \right\}_{m = 1}^M,\widehat{\bf{h}}_u^{\left( k \right)}} \right) \\ 
= \arg \mathop {\max }\limits_{\left( {{\varepsilon _u},\left\{ {{\theta _{u,m}}} \right\}_{m = 1}^M,{{\bf{h}}_u}} \right)} \log p\left( {\widehat{\bf{r}}_u^{\left( k \right)}\left| {{\varepsilon _u},\left\{ {{\theta _{u,m}}} \right\}_{m = 1}^M,{{\bf{h}}_u}} \right.} \right) \\ 
= \arg \mathop {\max }\limits_{\left( {{\varepsilon _u},\left\{ {{\theta _{u,m}}} \right\}_{m = 1}^M,{{\bf{h}}_u}} \right)}  \\ 
\;\;\;\;\;\;\;\;\;\;\;\;\;\;\;\;\;\;   \left\{ {\log \sum\limits_{{{\bf{X}}_u}} {p\left( {\widehat{\bf{r}}_u^{\left( k \right)},{{\bf{X}}_u}\left| {{\varepsilon _u},\left\{ {{\theta _{u,m}}} \right\}_{m = 1}^M,{{\bf{h}}_u},{C_u}} \right.} \right)} } \right\} \\ 
\end{array}
\end{equation}
Note that problem (\ref{ecmprob}) is different from problem  (\ref{subprob}) , where all variables $\left\{ {{\varepsilon _u},\left\{ {{\theta _{u,m}}} \right\}_{m = 1}^M,{{\bf{h}}_u},{{\bf{X}}_u}} \right\}$  (including ${{{\bf{X}}_u}}$) are treated as parameters to be estimated.

After ECM finds $\left\{ {\widehat\varepsilon _u^{\left( k \right)},\left\{ {\widehat\theta _{u,m}^{\left( k \right)}} \right\}_{m = 1}^M,\widehat{\bf{h}}_u^{\left( k \right)}} \right\}$ in (\ref{ecmprob}),  we then set $\left\{ {{\varepsilon _u},\left\{ {{\theta _{u,m}}} \right\}_{m = 1}^M,{{\bf{h}}_u}} \right\} = \left\{ {\widehat\varepsilon _u^{\left( k \right)},\left\{ {\widehat\theta _{u,m}^{\left( k \right)}} \right\}_{m = 1}^M,\widehat{\bf{h}}_u^{\left( k \right)}} \right\}$ in the signal model (\ref{sm2}) and then perform sum-product channel decoding to find the a posteriori probabilities (APPs) of the coded symbol $p\left( {{X_{u,m,i}}\left| {\widehat{\bf{r}}_u^{\left( k \right)},\widehat\varepsilon _u^{\left( k \right)},\left\{ {\widehat\theta _{u,m}^{\left( k \right)}} \right\}_{m = 1}^M,\widehat{\bf{h}}_u^{\left( k \right)},{C_u}} \right.} \right)$
for all $m$ and $i$.  For example, if the convolution code is used, the corresponding  channel decoding algorithm is the BCJR algorithm \cite{bcjrtit}. The estimates for the symbols in ${{\bf{X}}_u}$ are obtained by making hard decisions based on the symbol-wise APPs:
$$\begin{array}{l}
\hat X_{u,m,i}^{\left( k \right)} \\ 
= \arg \mathop {\max }\limits_{{X_{u,m,i}}} p\left( {{X_{u,m,i}}\left| {\widehat{\bf{r}}_u^{\left( k \right)},\hat \varepsilon _u^{\left( k \right)},\left\{ {\hat \theta _{u,m}^{\left( k \right)}} \right\}_{m = 1}^M,\widehat{\bf{h}}_u^{\left( k \right)},{C_u}} \right.} \right) \\ 
\end{array}$$
We stack all the symbol estimates $\widehat X_{u,m,i}^{\left( k \right)}$ for all $m$ and  $i$ into a vector $\widehat{\bf{X}}_u^{\left( k \right)}$, and $\widehat{\bf{X}}_u^{\left( k \right)}$ is treated as the estimate for ${{\bf{X}}_u}$. 

Finally, the parameter estimates $\left\{ {\widehat\varepsilon _u^{\left( k \right)},\left\{ {\widehat\theta _{u,m}^{\left( k \right)}} \right\}_{m = 1}^M,\widehat{\bf{h}}_u^{\left( k \right)}} \right\}$ by ECM together with the hidden data estimate $\widehat{\bf{X}}_u^{\left( k \right)}$ by channel decoding are treated as the overall solution for $\left\{ {\widehat\varepsilon _u^{\left( k \right)},\left\{ {\widehat\theta _{u,m}^{\left( k \right)}} \right\}_{m = 1}^M,\widehat{\bf{h}}_u^{\left( k \right)},\widehat{\bf{X}}_u^{\left( k \right)}} \right\}$. We will see that the ECM and channel decoding can be integrated into one framework to assist each other for finding $\left\{ {\widehat\varepsilon _u^{\left( k \right)},\left\{ {\widehat\theta _{u,m}^{\left( k \right)}} \right\}_{m = 1}^M,\widehat{\bf{h}}_u^{\left( k \right)}} \right\}$ and $\widehat{\bf{X}}_u^{\left( k \right)}$. Here, we emphasize that $\widehat{\bf{X}}_u^{\left( k \right)}$ are the decoded soft information (APPs). We will show in the following how to use the APPs to refine the estimates of the parameters through the ECM iterations.

Within the $k^{th}$ iteration of SAGE, the ECM algorithm for solving each sub-problem in (\ref{subprob}) consists of $Z$ iterations. We collect the parameters into a set ${\Omega _u} \buildrel \Delta \over = \left\{ {{\varepsilon _u},\left\{ {{\theta _{u,m}}} \right\}_{m = 1}^M,{{\bf{h}}_u}} \right\}$, and denote the updated parameter estimates after the $z^{th}$ ECM iteration within the $k^{th}$ SAGE iteration by $\widehat\Omega _u^{\left( {k,z} \right)} = \left\{ {\widehat\varepsilon _u^{\left( {k,z} \right)},\left\{ {\widehat\theta _{u,m}^{\left( {k,z} \right)}} \right\}_{m = 1}^M,\widehat{\bf{h}}_u^{\left( {k,z} \right)}} \right\}$, $z = 1,2, \cdots ,Z$. The $z^{th}$  iteration of ECM within the $k^{th}$  SAGE iteration consists of an E-step and an M-step, as follows:
\\

\noindent \textbf{E-step of ECM}: 

\noindent Given the parameter estimates from the previous ECM iteration $\widehat\Omega _u^{\left( {k,z - 1} \right)}$, the E-step of ECM aims to compute the conditional expectation defined by \cite{mclachlan2007algorithm}.
\begin{equation}\label{qf1}
\begin{array}{*{20}{l}}
{Q\left( {{\Omega _u}\left| {\hat \Omega _u^{\left( {k,z - 1} \right)}} \right.} \right)}  \\
{ = \sum\limits_{{{\bf{X}}_u}} {p\left( {{{\bf{X}}_u}\left| {\widehat{\bf{r}}_u^{\left( k \right)},\hat \Omega _u^{\left( {k,z - 1} \right)},{C_u}} \right.} \right)\log p\left( {\widehat{\bf{r}}_u^{\left( k \right)}\left| {{\Omega _u},{{\bf{X}}_u}} \right.} \right)} }  \\
\end{array}
\end{equation}       
This conditional expectation is called the Q function in the EM literature. We remark that the decoded soft information of the transmitted symbols, if obtainable, is used to assist the estimate of the parameters. With this setup, the data symbols serve as the hidden data in the ECM framework. Since the expectation (the summation) in (\ref{qf1}) is taken over the data symbols in the frequency domain, we might be tempted compute the above Q function in the frequency domain. However, we will see that this does not work. With a new approach, we can compute the Q function in the time domain. 
\\

\noindent  \emph{How to decompose Q function in time domain}

A brute-force attack that attempts to compute the Q function in (\ref{qf1}) directly will meet the following obstacles: (i) the ensemble of codewords ${{\bf{X}}_u}$ to be summed over in (\ref{qf1}) is very large (in fact, exponentially large in terms of the number of symbols in ${{\bf{X}}_u}$); (ii) barring exhaustive enumeration, there is no known decoding algorithm that can give $p\left( {{{\bf{X}}_u}\left| {\widehat{\bf{r}}_u^{\left( k \right)},\widehat\Omega _u^{\left( {k,z - 1} \right)},{C_u}} \right.} \right)$ for all these codewords; (iii) there is no concise expression for $p\left( {\widehat{\bf{r}}_u^{\left( k \right)}\left| {{\Omega _u},{{\bf{X}}_u}} \right.} \right)$ due to the coupled terms brought about by the ICI effect.  Ref. \cite{pun2007iterative} did not take these obstacles into account, since it investigated uncoded systems. In order to obtain a practical solution for (\ref{qf1}), in this work, we transform the computation process to the time domain, where the Q function can be decomposed into many sample-wise factors.

We define the frequency-domain compound symbol of the subcarrier symbol and the channel gain as
$${Y_{u,m,i}} \buildrel \Delta \over = {X_{u,m,i}}{H_{u,i}}$$
Given the channel gain ${H_{u,i}}$, ${Y_{u,m,i}} \in \left\{ { \pm {H_{u,i}}} \right\}$ when BPSK is adopted. The vector of frequency-domain compound symbols for the $m^{th}$ OFDM block is 
$$
\begin{array}{l}
{{\bf{Y}}_{u,m}} \buildrel \Delta \over = {\left[ {{Y_{u,m,1}}{Y_{u,m,2}} \cdots {Y_{u,m,N}}} \right]^T} 
 = diag\left\{ {{{\bf{X}}_{u,m}}} \right\}{{\bf{H}}_u}; \\ 
\end{array}
$$
the corresponding vector of time-domain compound samples is ${{\bf{y}}_{u,m}} \buildrel \Delta \over = {\left[ {{y_{u,m,1}}{y_{u,m,2}} \cdots {y_{u,m,N}}} \right]^T} = {{\bf{F}}^H}{{\bf{Y}}_{u,m}}$; and the vector of frequency-domain compound symbols (time-domain compound samples) for the whole frame is ${{\bf{Y}}_u} \buildrel \Delta \over = {\left[ {{\bf{Y}}_{u,1}^T{\bf{Y}}_{u,2}^T \cdots {\bf{Y}}_{u,M}^T} \right]^T}$
(${{\bf{y}}_u} \buildrel \Delta \over = {\left[ {{\bf{y}}_{u,1}^T{\bf{y}}_{u,2}^T \cdots {\bf{y}}_{u,M}^T} \right]^T}$). Bear in mind that given the channel gains ${{\bf{H}}_u}$, the knowledge of the compound symbols ${{\bf{Y}}_u}$
is equivalent to the knowledge of  the transmit symbols ${{\bf{X}}_u}$. We can therefore rewrite the Q function in (\ref{qf1}) as 
\begin{equation}\label{qf2}
	\begin{array}{l}
		Q\left( {{\Omega _u}\left| {\widehat\Omega _u^{\left( {k,z - 1} \right)}} \right.} \right) \\ 
		= \sum\limits_{{{\bf{Y}}_u}} {\left\{ {p\left( {{{\bf{Y}}_u}\left| {\widehat{\bf{r}}_u^{\left( k \right)},\widehat\Omega _u^{\left( {k,z - 1} \right)},{C_u}} \right.} \right)\log p\left( {\widehat{\bf{r}}_u^{\left( k \right)}\left| {{\Omega _u},{{\bf{Y}}_u}} \right.} \right)} \right\}} {\rm{ }} \\ 
		= \sum\limits_{{{\bf{y}}_u}} {\left\{ {p\left( {{{\bf{y}}_u}\left| {\widehat{\bf{r}}_u^{\left( k \right)},\widehat\Omega _u^{\left( {k,z - 1} \right)},{C_u}} \right.} \right)\log p\left( {\widehat{\bf{r}}_u^{\left( k \right)}\left| {{\Omega _u},{{\bf{y}}_u}} \right.} \right)} \right\}}  \\ 
	\end{array}
\end{equation}
where the second equality is due to that the mapping between ${{\bf{Y}}_u}$ and ${{\bf{y}}_u}$ is a one-to-one correspondence.

Next, we decompose the probability functions in (\ref{qf2}) to compute the time-domain Q function. We denote the $i^{th}$ element of $\widehat{\bf{r}}_{u,m}^{\left( k \right)}$ (the estimate for the $i^{th}$  sample of the $m^{th}$ OFDM block of user $u$) by $\widehat r_{u,m,i}^{\left( k \right)}$. Since CFO only introduces a linear phase drift in the time-domain signals (and not ICI), we can immediately decouple the components $\widehat r_{u,m,i}^{\left( k \right)}$  in $\widehat{\bf{r}}_u^{\left( k \right)}$ and decompose $p\left( {\widehat{\bf{r}}_u^{\left( k \right)}\left| {{\Omega _u},{{\bf{y}}_u}} \right.} \right)$ as 
\begin{equation}\label{pf1}
	p\left( {\widehat{\bf{r}}_u^{\left( k \right)}\left| {{\Omega _u},{{\bf{y}}_u}} \right.} \right) = \prod\limits_{m = 1}^M {\prod\limits_{i = 1}^N {p\left( {\widehat r_{u,m,i}^{\left( k \right)}\left| {{\Omega _u},{y_{u,m,i}}} \right.} \right)} } 
\end{equation}                           
where the sample-wise factor is given by
\begin{equation}
\begin{array}{l}
p\left( {\widehat r_{u,m,i}^{\left( k \right)}\left| {{\Omega _u},{y_{u,m,i}}} \right.} \right) \\ 
= {\cal{CN}} \left( {\widehat r_{u,m,i}^{\left( k \right)}:{e^{j{\theta _{u,m}}}}{e^{j2\pi {\varepsilon _u}{{\left( {i - 1} \right)} \mathord{\left/
					{\vphantom {{\left( {i - 1} \right)} N}} \right.
					\kern-\nulldelimiterspace} N}}}{y_{u,m,i}},\sigma _n^2} \right) \\ 
\end{array}
\end{equation}
for all $m$ and $i$. Substituting (\ref{pf1}) into (\ref{qf2}), we obtain 
\begin{equation}
\begin{array}{l}
Q\left( {{\Omega _u}\left| {\widehat\Omega _u^{\left( {k,z - 1} \right)}} \right.} \right){\rm{  = }}\sum\limits_{m = 1}^M {\sum\limits_{i = 1}^N {\sum\limits_{{y_{u,m,i}}} {\log p\left( {\widehat r_{u,m,i}^{\left( k \right)}\left| {{\Omega _u},{y_{u,m,i}}} \right.} \right)} } }  \\ 
\;\;\;\;\;\;\;\;\;\;\;\;\;\;\;\;\;\;\;\;\;\;\;\;\;\;\;\;  \times \underbrace {\sum\limits_{\left\{ {{{\bf{y}}_u}: \sim {y_{u,m,i}}} \right\}} {p\left( {{{\bf{y}}_u}\left| {\widehat{\bf{r}}_u^{\left( k \right)},\widehat\Omega _u^{\left( {k,z - 1} \right)},{C_u}} \right.} \right)} }_{ \buildrel \Delta \over = p\left( {{y_{u,m,i}}\left| {\widehat{\bf{r}}_u^{\left( k \right)},\widehat\Omega _u^{\left( {k,z - 1} \right)},{C_u}} \right.} \right)} \\ 
\end{array}
\end{equation}
where $\left\{ {{{\bf{y}}_u}: \sim {y_{u,m,i}}} \right\}$ is the set that contains the all elements in vector ${{\bf{y}}_u}$ except ${y_{u,m,i}}$, and   $p\left( {{y_{u,m,i}}\left| {\widehat{\bf{r}}_u^{\left( k \right)},\widehat\Omega _u^{\left( {k,z - 1} \right)},{C_u}} \right.} \right)$ is the sample-wise APP. We define the sample-wise Q function as 
\begin{equation}\label{sqf}
	\begin{array}{l}
		{Q_{m,i}}\left( {{\Omega _u}\left| {\widehat\Omega _u^{\left( {k,z - 1} \right)}} \right.} \right) 
		\buildrel \Delta \over  = \\
				 \sum\limits_{{y_{u,m,i}}} {\log p\left( {\widehat r_{u,m,i}^{\left( k \right)}\left| {{\Omega _u},{y_{u,m,i}}} \right.} \right)} p\left( {{y_{u,m,i}}\left| {\widehat{\bf{r}}_u^{\left( k \right)},\widehat\Omega _u^{\left( {k,z - 1} \right)},{C_u}} \right.} \right) \\ 
	\end{array}
\end{equation}
Finally, the overall Q function is the sum of sample-wise Q functions:
\begin{equation}
	Q\left( {{\Omega _u}\left| {\widehat\Omega _u^{\left( {k,z - 1} \right)}} \right.} \right){\rm{  }} = \sum\limits_{m = 1}^M {\sum\limits_{i = 1}^N {{Q_{m,i}}\left( {{\Omega _u}\left| {\widehat\Omega _u^{\left( {k,z - 1} \right)}} \right.} \right){\rm{ }}} } 
\end{equation}
As can be seen from the above, unlike the frequency-domain Q function, the time-domain Q function can be computed on a sample-by-sample basis. This greatly reduces the computational complexity. 
\\

\noindent \emph{How to transform APPs from frequency domain to time domain}

The next question is how to obtain the APPs of the time-domain samples $\left\{ {p\left( {{y_{u,m,i}}\left| {\widehat{\bf{r}}_u^{\left( k \right)},\widehat\Omega _u^{\left( {k,z - 1} \right)},{C_u}} \right.} \right)} \right\}$ in (\ref{sqf}).  Now, the symbols are channel-coded and transmitted in the frequency domain; after DFT the receiver performs channel decoding to find the APPs of the transmit symbols $\left\{ {p\left( {{X_{u,m,i}}\left| {\widehat{\bf{r}}_u^{\left( k \right)},\widehat\Omega _u^{\left( {k,z - 1} \right)},{C_u}} \right.} \right)} \right\}$. Given the estimated channel gains  $\widehat{\bf{H}}_u^{\left( {k,z - 1} \right)} \buildrel \Delta \over = {\left[ {\widehat H_{u,1}^{\left( {k,z - 1} \right)}\widehat H_{u,2}^{\left( {k,z - 1} \right)} \cdots \widehat H_{u,N}^{\left( {k,z - 1} \right)}} \right]^T}{\rm{ = }}{\bf{F}}\widehat{\bf{h}}_u^{\left( {k,z - 1} \right)}$, this is equivalent to finding the APPs of the frequency-domain compound symbols, i.e., 
$
p\left( {{X_{u,m,i}}\left| {\widehat{\bf{r}}_u^{\left( k \right)},\widehat\Omega _u^{\left( {k,z - 1} \right)},{C_u}} \right.} \right) = p\left( {{Y_{u,m,i}}\left| {\widehat{\bf{r}}_u^{\left( k \right)},\widehat\Omega _u^{\left( {k,z - 1} \right)},{C_u}} \right.} \right)
$
where ${Y_{u,m,i}} = {X_{u,m,i}}\widehat H_{u,i}^{\left( {k,z - 1} \right)}$ for all $m$ and $i$.  The algorithm to achieve this decoding objective is the general sum-product algorithm \cite{wiberg1996codes, kschischang2001factor}.  

There is no known technique, however, for decoding the time-domain samples $\left\{ {{y_{u,m,i}}} \right\}$. In this work, we introduce a new concept called “soft IDFT” to transform the APPs obtained by the sum-product decoding algorithm from the frequency domain to the time domain. 
At first sight, we might be tempted to obtain $p\left( {{{\bf{Y}}_{u,m}}\left| {\widehat{\bf{r}}_u^{\left( k \right)},} \right.} \right.$ $\left. {\widehat\Omega _u^{\left( {k,z - 1} \right)},{C_u}} \right)$ from $p\left( {{{\bf{Y}}_{u,m}}\left| {\widehat{\bf{r}}_u^{\left( k \right)},\widehat\Omega _u^{\left( {k,z - 1} \right)},{C_u}} \right.} \right)$ by considering the linear transformation between ${{\bf{y}}_{u,m}}$ and ${{\bf{Y}}_{u,m}}$, i.e.,  , ${{\bf{y}}_{u,m}} = {{\bf{F}}^H}{{\bf{Y}}_{u,m}}$ and then deriving the APP of sample ${y_{u,m,i}}$ by marginalizing out all other samples over $p\left( {{{\bf{y}}_{u,m}}\left| {\widehat{\bf{r}}_u^{\left( k \right)},\widehat\Omega _u^{\left( {k,z - 1} \right)},{C_u}} \right.} \right)$.  There are two obstacles to this approach. First, it is hard to obtain the APP of the vector of symbols $p\left( {{{\bf{Y}}_{u,m}}\left| {\widehat{\bf{r}}_u^{\left( k \right)},\widehat\Omega _u^{\left( {k,z - 1} \right)},{C_u}} \right.} \right)$  (the sum-product algorithm, for example, does not give this APP).  Second, the aforementioned marginalization will introduce intractable complexity that is in the exponential order of the DFT size $N$.  Here, we employ a Gaussian message passing \cite{loeliger2007factor} approach to solve the soft IDFT problem approximately.

We approximate the APPs of the frequency-domain symbols as independent Gaussian distributions. The independence can be achieved by the operation of interleaving in the transmitter. The Gaussianity is an assumption made to simplify computation.\footnote{Without the Gaussian assumption, we can solve the soft IDFT problem on a factor graph using the general message passing algorithm. We first construct the factor graph by considering the butterfly graph of FFT. With $\left\{ {p\left( {{Y_{u,m,i}}\left| {\widehat{\bf{r}}_u^{\left( k \right)},\widehat\Omega _u^{\left( {k,z - 1} \right)},{C_u}} \right.} \right)} \right\}_{i = 1}^N$ as the input messages, we use sum-product rule to compute the output messages $\left\{ {p\left( {{y_{u,m,i}}\left| {\widehat{\bf{r}}_u^{\left( k \right)},\widehat\Omega _u^{\left( {k,z - 1} \right)},C} \right.} \right)} \right\}_{i = 1}^N$ exactly. However, the number of messages on the graph increases in an exponential order as we progress from the input to the output of the IDFT factor graph (i.e., unlike the binary ${Y_{u,m,i}}$, ${y_{u,m,i}}$ has $2^N$ possible values.).  We will treat this problem in our future study.} The approximate APP of ${Y_{u,m,i}}$ is given by 
\begin{equation}
\begin{array}{l}
\widetilde p\left( {{Y_{u,m,i}}\left| {\widehat{\bf{r}}_u^{\left( k \right)},\hat \Omega _u^{\left( {k,z - 1} \right)},{C_u}} \right.} \right) \\ 
\;\;\;\;\;\;\;\;\;\;\;\;\;\;\;\;\;\;\;\;\;\; = {\cal{CN}}\left( {{Y_{u,m,i}}:{m_{{Y_{u,m,i}}}},\sigma _{{Y_{u,m,i}}}^2} \right) \\ 
\end{array}
\end{equation}
with mean and variance 
$${m_{{Y_{u,m,i}}}} = \sum\nolimits_{{Y_{u,m,i}}} {{Y_{u,m,i}}p\left( {{Y_{u,m,i}}\left| {\widehat{\bf{r}}_u^{\left( k \right)},\widehat\Omega _u^{\left( {k,z - 1} \right)},{C_u}} \right.} \right)} $$
$$\begin{array}{*{20}{l}}
{\sigma _{{Y_{u,m,i}}}^2 = }  \\
{{\sum _{{Y_{u,m,i}}}}{{\left\| {{Y_{u,m,i}} - {m_{{Y_{u,m,i}}}}} \right\|}^2}p\left( {{Y_{u,m,i}}\left| {\widehat{\bf{r}}_u^{\left( k \right)},\hat \Omega _u^{\left( {k,z - 1} \right)},{C_u}} \right.} \right)}  \\
\end{array}$$
which are computed using the APPs delivered from the sum-product channel decoding. Then, with the independence assumption, we can write the APP of ${{\bf{Y}}_{u,m}}$ as 
\begin{equation}
\begin{array}{l}
p\left( {{{\bf{Y}}_{u,m}}\left| {\widehat{\bf{r}}_u^{\left( k \right)},\hat \Omega _u^{\left( {k,z - 1} \right)},{C_u}} \right.} \right) \\ 
\;\;\;\;\;\;\;\;\;\;\;\;\;\;\;\;\;\;\;\;\; = \prod\limits_{i = 1}^N {\widetilde p\left( {{Y_{u,m,i}}\left| {\widehat{\bf{r}}_u^{\left( k \right)},\hat \Omega _u^{\left( {k,z - 1} \right)},{C_u}} \right.} \right)}  \\ 
\;\;\;\;\;\;\;\;\;\;\;\;\;\;\;\;\;\;\;\;\; = {\cal CN}\left( {{{\bf{Y}}_{u,m}}:{{\bf{m}}_{{{\bf{Y}}_{u,m}}}},{{\bf{C}}_{{{\bf{Y}}_{u,m}}}}} \right) \\ 
\end{array}
\end{equation}
with mean vector and covariance matrix
$$\begin{array}{l}
{{\bf{m}}_{{{\bf{Y}}_{u,m}}}} \buildrel \Delta \over = {\left[ {{m_{{Y_{u,m,1}}}}{m_{{Y_{u,m,2}}}} \cdots {m_{{Y_{u,m,N}}}}} \right]^T} \\ 
{{\bf{C}}_{{{\bf{Y}}_{u,m}}}} = diag\left( {\left[ {\sigma _{{Y_{u,m,1}}}^2\sigma _{{Y_{u,m,2}}}^2 \cdots \sigma _{{Y_{u,m,N}}}^2} \right]} \right) \\ 
\end{array}$$
Finally, since ${{\bf{y}}_{u,m}} = {{\bf{F}}^H}{{\bf{Y}}_{u,m}}$ and ${{\bf{F}}^H}$ is unitary and ${{\bf{Y}}_{u,m}}$ is assumed to be Gaussian distributed, the APP of ${{\bf{y}}_{u,m}}$ is given by
\begin{equation}
	p\left( {{{\bf{y}}_{u,m}}\left| {\widehat{\bf{r}}_u^{\left( k \right)},\widehat\Omega _u^{\left( {k,z - 1} \right)},{C_u}} \right.} \right) = {\cal{CN}}\left( {{{\bf{y}}_{u,m}}:{{\bf{m}}_{{{\bf{y}}_{u,m}}}},{{\bf{C}}_{{{\bf{y}}_{u,m}}}}} \right)
\end{equation}
with mean vector and covariance matrix
$$\begin{array}{l}
{{\bf{m}}_{{{\bf{y}}_{u,m}}}} = {{\bf{F}}^H}{{\bf{m}}_{{{\bf{Y}}_{u,m}}}} \\ 
{{\bf{C}}_{{{\bf{y}}_{u,m}}}} = {{\bf{F}}^H}{{\bf{C}}_{{{\bf{Y}}_{u,m}}}}{\bf{F}}. \\ 
\end{array}$$
A nice feature of Gaussian distributions is that every marginal distribution of a joint Gaussian distribution is itself a Gaussian distribution \cite{ahrendt2005multivariate}, and the APP of ${y_{u,m,i}}$ is immediately given by 
\begin{equation}\label{gauss}
	p\left( {{y_{u,m,i}}\left| {\widehat{\bf{r}}_u^{\left( k \right)},\widehat\Omega _u^{\left( {k,z - 1} \right)},C} \right.} \right) = {\cal{CN}}\left( {{y_{u,m,i}}:{m_{{y_{u,m,i}}}},\sigma _{{y_{u,m,i}}}^2} \right)
\end{equation}
with mean and variance 
$$
\begin{array}{l}
{m_{{y_{u,m,i}}}} = {\left[ {{{\bf{m}}_{{{\bf{y}}_{u,m}}}}} \right]_i} = \frac{1}{{\sqrt N }}\sum\limits_{j = 1}^{N } {{{\mathop{\rm e}\nolimits} ^{j2\pi {{(i-1)(j-1)} \mathord{\left/
					{\vphantom {{ij} N}} \right.
					\kern-\nulldelimiterspace} N}}}} {m_{{Y_{u,m,j}}}} \\ 
\sigma _{{y_{u,m,i}}}^2 = {\left[ {{{\bf{C}}_{{{\bf{y}}_{u,m}}}}} \right]_{i,i}}. \\ 
\end{array}
$$
We will see later that when we use the Gaussian form of the APP shown in (\ref{gauss}) for computing the Q function, only the mean ${m_{{y_{u,m,i}}}}$ has impact on the actual form of the Q function and the variance $\sigma _{{y_{u,m,i}}}^2$ can be dropped. Since the mean of ${y_{u,m,i}}$ is easy to compute because it is a linear combination of the means of $\left\{ {{Y_{u,m,i}}} \right\}_{i = 1}^N$, the complexity of soft IDFT is reduced to the linear order of $N$ by Gaussian message passing. 
\\

\noindent \emph{How to obtain frequency-domain APPs}

We have considered how to obtain the time-domain sample-wise APPs from the frequency-domain symbol-wise APPs, given that the symbol-wise APPs $\left\{ {p\left( {{X_{u,m,i}}\left( {{Y_{u,m,i}}} \right)\left| {\widehat{\bf{r}}_u^{\left( k \right)},\widehat\Omega _u^{\left( {k,z - 1} \right)},{C_u}} \right.} \right)} \right\}$ are already computed by the sum-product channel decoding algorithm in the frequency domain. 

To compute the symbol-wise APPs in the frequency domain, we need the estimates for the frequency-domain signals of user $u$.  We have already computed the estimates for the time-domain signals $\left\{ {\widehat{\bf{r}}_{u,m}^{\left( k \right)}} \right\}$ as in (\ref{e-sage}). Before transforming $\left\{ {\widehat{\bf{r}}_{u,m}^{\left( k \right)}} \right\}$ into the frequency domain, we compensate for the CFO using the CFO estimate $\widehat\varepsilon _u^{\left( {k,z - 1} \right)}$ from the last iteration. Then, we perform DFT on the compensated estimates for the time-domain signals 
$$\widehat{\bf{R}}_{u,m}^{\left( k \right)} = {\bf{F}}\left( {{\bf{\Gamma }}\left( { - \widehat\varepsilon _u^{\left( {k,z - 1} \right)}} \right)\widehat{\bf{r}}_{u,m}^{\left( k \right)}} \right)$$                                                 
for $m = 1,2, \cdots ,M$. After that, we can obtain the evidence information from $\left\{ {\widehat{\bf{R}}_{u,m}^{\left( k \right)}} \right\}$
\begin{equation}\label{evidence}
\begin{array}{l}
p\left( {\hat R_{u,m,i}^{\left( k \right)}\left| {{X_{u,m,i}},\hat \Omega _u^{\left( {k,z - 1} \right)}} \right.} \right) \\ 
= {\cal CN} \left( {\hat R_{u,m,i}^{\left( k \right)}:{e^{j\hat \theta _{u,m}^{\left( {k,z - 1} \right)}}}\hat H_{u,i}^{\left( {k,z - 1} \right)}{X_{u,m,i}},\sigma _{IN}^2} \right) \\ 
\end{array}
\end{equation}
where $i = 1,2, \cdots ,N$, $m = 1,2, \cdots ,M$, $\widehat R_{u,m,i}^{\left( k \right)}$ is the $i^{th}$ element of $\widehat{\bf{R}}_{u,m}^{\left( k \right)}$, and $\sigma _{IN}^2$ is the variance of the residual interference plus noise. The residual interference remaining in $\left\{ {\widehat{\bf{R}}_{u,m}^{\left( k \right)}} \right\}$ includes the residual inter-carrier interference and multiple user interference. The variance $\sigma _{IN}^2$ is an unknown variable whose value is changing over the iterations. Before channel decoding, we employ a simple method for estimating $\sigma _{IN}^2$ in each iteration 
\begin{equation}
	\widehat{\sigma _{IN}^2} = \frac{1}{{NM}}\sum\limits_{m = 1}^M {\sum\limits_{i = 1}^N {\left( {{{\left\| {\widehat R_{u,m,i}^{\left( k \right)}} \right\|}^2} - {{\left\| {\widehat H_{u,i}^{\left( {k,z - 1} \right)}} \right\|}^2}} \right)} } 
\end{equation}
The estimate $\widehat{\sigma _{IN}^2}$ is used to replace $\sigma _{IN}^2$ in (\ref{evidence}) when we compute the evidence information.  

The evidence information computed above is used to initialize the standard sum-product algorithm for channel decoding. We can derive the sum-product channel decoding algorithm as a message passing algorithm on the factor graph that models the encoding constraint $C_u$. Readers familiar with the sum-product algorithm and factor graphs \cite{kschischang2001factor} can readily complete this task; we omit the details here. 
\\

\noindent \emph{Ultimate Form for Q Function}

Finally, using the Gaussian expression of the sample-wise APP shown in (\ref{gauss}), we can simplify the sample-wise Q function (\ref{sqf}) into  a compact form (the derivation is given in Appendix A):
\begin{equation}\label{sqf2}
\begin{array}{l}
{Q_{m,i}}\left( {{\Omega _u}\left| {\widehat\Omega _u^{\left( {k,z - 1} \right)}} \right.} \right) \\ 
\propto  - {\left\| {\widehat r_{u,m,i}^{\left( k \right)} - {e^{j{\theta _{u,m}}}}{e^{j2\pi {\varepsilon _u}{{\left( {i - 1} \right)} \mathord{\left/
						{\vphantom {{\left( {i - 1} \right)} N}} \right.
						\kern-\nulldelimiterspace} N}}}{m_{{y_{u,m,i}}}}} \right\|^2} \\ 
\end{array}
\end{equation}
From (\ref{sqf2}), we note that, due to the Gaussian expression for the APP of ${y_{u,m,i}}$ in (\ref{gauss}), only the mean of the time-domain sample ${y_{u,m,i}}$  appears in the sample-wise Q function. This greatly reduces  the complexities involved in the computation of the Q function. With  (\ref{sqf2}), the overall Q function can now be written as
\begin{equation}\label{overallqf}
\begin{array}{l}
Q\left( {{\Omega _u}\left| {\widehat\Omega _u^{\left( {k,z - 1} \right)}} \right.} \right) \\ 
\propto  - \sum\limits_{m = 1}^M {\sum\limits_{i = 1}^N {{{\left\| {\widehat r_{u,m,i}^{\left( k \right)} - {e^{j{\theta _{u,m}}}}{e^{j2\pi {\varepsilon _u}{{\left( {i - 1} \right)} \mathord{\left/
									{\vphantom {{\left( {i - 1} \right)} N}} \right.
									\kern-\nulldelimiterspace} N}}}{m_{{y_{u,m,i}}}}} \right\|}^2}{\rm{ }}} }  \\ 
\propto  - \sum\limits_{m = 1}^M {{{\left\| {\widehat{\bf{r}}_{u,m}^{\left( k \right)} - {e^{j{\theta _{u,m}}}}{\bf{\Gamma }}\left( {{\varepsilon _u}} \right){{\bf{m}}_{{{\bf{y}}_{u,m}}}}} \right\|}^2}} . \\ 
\end{array}
\end{equation}
We denote the mean vector of the symbol vector ${{\bf{X}}_{u,m}}$ by ${{\bf{m}}_{{{\bf{X}}_{u,m}}}} = {\left[ {{m_{{X_{u,m,1}}}}{m_{{X_{u,m,2}}}} \cdots {m_{{X_{u,m,N}}}}} \right]^T}$, whose $i^{th}$ element is computed using the symbol-wise APP
\begin{equation}
	{m_{{X_{u,m,i}}}} = \sum\limits_{{X_{u,m,i}}} {{X_{u,m,i}}p\left( {{X_{u,m,i}}\left| {\widehat{\bf{r}}_u^{\left( k \right)},\widehat\Omega _u^{\left( {k,z - 1} \right)},{C_u}} \right.} \right)} 
\end{equation}
With the above notations, we have the following relationship: 
\begin{equation}\label{meanvector}
	{{\bf{m}}_{{{\bf{y}}_{u,m}}}} = {{\bf{F}}^H}{{\bf{m}}_{{{\bf{Y}}_{u,m}}}} = {{\bf{F}}^H}{\bf{D}}\left( {{{\bf{m}}_{{{\bf{X}}_{u,m}}}}} \right){\bf{F}}{{\bf{h}}_u}
\end{equation}
Substituting (\ref{meanvector}) into (\ref{overallqf}) gives the ultimate form of the Q function:
\begin{equation}\label{finalqf}
\begin{array}{l}
Q\left( {{\Omega _u}\left| {\widehat\Omega _u^{\left( {k,z - 1} \right)}} \right.} \right) \\ 
\propto  - \sum\limits_{m = 1}^M {{{\left\| {\widehat{\bf{r}}_{u,m}^{\left( k \right)} - {e^{j{\theta _{u,m}}}}{\bf{\Gamma }}\left( {{\varepsilon _u}} \right){{\bf{F}}^H}{\bf{D}}\left( {{{\bf{m}}_{{{\bf{X}}_{u,m}}}}} \right){\bf{F}}{{\bf{h}}_u}} \right\|}^2}}  \\ 
\end{array}
\end{equation}
So far, we have finished the E-step of ECM.  We next turn to the M-step of ECM. 
\\

\noindent \textbf{M-step of ECM}:

The M-step of ECM updates the new parameter estimates $\widehat\Omega _u^{\left( {k,z} \right)} = \left\{ {\widehat\varepsilon _u^{\left( {k,z} \right)},\left\{ {\widehat\theta _{u,m}^{\left( {k,z} \right)}} \right\}_{m = 1}^M,\widehat{\bf{h}}_u^{\left( {k,z} \right)}} \right\}$ by maximizing the Q function in (\ref{finalqf}). ECM breaks the maximization procedure of the $z^{th}$ iteration into three stages, where the $t^{th}$ stage updates the parameters from $\widehat\Omega _u^{\left( {k,z - 1 + \frac{{t - 1}}{3}} \right)}$ to $\widehat\Omega _u^{\left( {k,z - 1 + \frac{t}{3}} \right)}$, $t = 1,2,3$, as follows:
\\

\noindent \emph{The First Stage --- CFO Estimation}

The first stage updates the CFO with the phase drifts and the channel gains fixed to their last estimates. The new set of parameter estimates after the first stage is $\widehat\Omega _u^{\left( {k,z - 1 + \frac{1}{3}} \right)} = \left\{ {\widehat\varepsilon _u^{\left( {k,z} \right)},\left\{ {\widehat\theta _{u,m}^{\left( {k,z - 1} \right)}} \right\}_{m = 1}^M,\widehat{\bf{h}}_u^{\left( {k,z - 1} \right)}} \right\}$. The new CFO estimate $\widehat\varepsilon _u^{\left( {k,z} \right)}$ is obtained by 
\begin{equation}\label{cfoe}
\begin{array}{l}
\widehat\varepsilon _u^{\left( {k,z} \right)} = \arg \mathop {\max }\limits_{{\varepsilon _u}} \left\{ \begin{array}{l}
\\ 
\\ 
\end{array} \right. \\ 
\left. { - \sum\limits_{m = 1}^M {{{\left\| {\widehat{\bf{r}}_{u,m}^{\left( k \right)} - {e^{j\widehat\theta _{u,m}^{\left( {k,z - 1} \right)}}}{\bf{\Gamma }}\left( {{\varepsilon _u}} \right){{\bf{F}}^H}{\bf{D}}\left( {{{\bf{m}}_{{{\bf{X}}_{u,m}}}}} \right){\bf{F}}\widehat{\bf{h}}_u^{\left( {k,z - 1} \right)}} \right\|}^2}} } \right\} \\ 
\end{array}
\end{equation}
where the objective function is the result of replacing the variables $\left\{ {{\theta _{u,m}}} \right\}_{m = 1}^M$ and ${{\bf{h}}_u}$ in the Q function (\ref{finalqf}) with the fixed values $\left\{ {\widehat\theta _{u,m}^{\left( {k,z - 1} \right)}} \right\}_{m = 1}^M$ and $\widehat{\bf{h}}_u^{\left( {k,z - 1} \right)}$.  

The exhaustive search method for solving (\ref{cfoe}) is computationally complex, since ${\varepsilon _u}$ is a continuous variable. To obtain a practical solution for (\ref{cfoe}), we could approximate ${\bf{\Gamma }}\left( {{\varepsilon _u}} \right)$ in (\ref{cfoe}) using its Taylor expansion around $\widehat\varepsilon _u^{\left( {k,z - 1} \right)}$ with terms above the second order truncated. We then differentiate the resulting objective function of (\ref{cfoe}) with respect to ${\varepsilon _u}$  and set the derivative to zero. Solving the equation yields a closed-form solution for (\ref{cfoe}):
\begin{equation}\label{cfoest}
	\begin{array}{l}
		\widehat\varepsilon _u^{\left( {k,z} \right)} = \widehat\varepsilon _u^{\left( {k,z - 1} \right)} 
		\\ + \frac{{Re\left\{ {\sum\limits_{m = 1}^M {{e^{j\widehat\theta _{u,m}^{\left( {k,z - 1} \right)}}}\widehat{\bf{r}}_{u,m}^{\left( k \right)H}{\bf{\Gamma '}}\left( {\widehat\varepsilon _u^{\left( {k,z - 1} \right)}} \right){{\bf{F}}^H}{\bf{D}}\left( {{{\bf{m}}_{{{\bf{X}}_{u,m}}}}} \right){\bf{F}}\widehat{\bf{h}}_u^{\left( {k,z - 1} \right)}} } \right\}}}{{Re\left\{ { - \sum\limits_{m = 1}^M {{e^{j\widehat\theta _{u,m}^{\left( {k,z - 1} \right)}}}\widehat{\bf{r}}_{u,m}^{\left( k \right)H}{\bf{\Gamma ''}}\left( {\widehat\varepsilon _u^{\left( {k,z - 1} \right)}} \right){{\bf{F}}^H}{\bf{D}}\left( {{{\bf{m}}_{{{\bf{X}}_{u,m}}}}} \right){\bf{F}}\widehat{\bf{h}}_u^{\left( {k,z - 1} \right)}} } \right\}}} \\ 
	\end{array}
\end{equation}
where 
$$\begin{array}{l}
{\bf{\Gamma '}}\left( {\widehat\varepsilon _u^{\left( {k,z - 1} \right)}} \right) = \left( {{{j2\pi } \mathord{\left/
			{\vphantom {{j2\pi } N}} \right.
			\kern-\nulldelimiterspace} N}} \right){\bf{\Psi \Gamma }}\left( {\widehat\varepsilon _u^{\left( {k,z - 1} \right)}} \right) \\ 
{\bf{\Gamma ''}}\left( {\widehat\varepsilon _u^{\left( {k,z - 1} \right)}} \right) =  - {\left( {{{2\pi } \mathord{\left/
				{\vphantom {{2\pi } N}} \right.
				\kern-\nulldelimiterspace} N}} \right)^2}{{\bf{\Psi }}^2}{\bf{\Gamma }}\left( {\widehat\varepsilon _u^{\left( {k,z - 1} \right)}} \right) \\ 
\end{array}
$$
with  ${\bf{\Psi }} = diag\left\{ {\left[ {0,1,2, \cdots ,N - 1} \right]} \right\}$, are the first and second derivates of ${\bf{\Gamma }}\left( {{\varepsilon _u}} \right)$ at the point $\widehat\varepsilon _u^{\left( {k,z - 1} \right)}$. The detailed derivation of (\ref{cfoest}) can be found in Appendix B. With the updated CFO estimate, we go to the second stage of the M-step.
\\

\noindent \emph{The Second Stage --- Phase Tracking}

The second stage updates the phase drifts with the CFO and the channel gains fixed to their last estimates. The new set of parameter estimates after the second stage is $\widehat\Omega _u^{\left( {k,z - 1 + \frac{2}{3}} \right)} = \left\{ {\widehat\varepsilon _u^{\left( {k,z} \right)},\left\{ {\widehat\theta _{u,m}^{\left( {k,z} \right)}} \right\}_{m = 1}^M,\widehat{\bf{h}}_u^{\left( {k,z - 1} \right)}} \right\}$, where the new phase estimates are given by 
\begin{equation}\label{phae}
\begin{array}{l}
 \left\{ {\hat \theta _{u,m}^{\left( {k,z} \right)}} \right\}_{m = 1}^M = \arg \mathop {\max }\limits_{\left\{ {{\theta _{u,m}}} \right\}_{m = 1}^M}  - \sum\limits_{m = 1}^M {\left\{ \begin{array}{l}
  \\ 
  \\ 
 \end{array} \right.}  \\ 
 \left. {{{\left\| {\widehat{\bf{r}}_{u,m}^{\left( k \right)} - {e^{j{\theta _{u,m}}}}\Gamma \left( {\hat \varepsilon _u^{\left( {k,z} \right)}} \right){{\bf{F}}^H}{\bf{D}}\left( {{{\bf{m}}_{{{\bf{X}}_{u,m}}}}} \right){\bf{F}}\widehat{\bf{h}}_u^{\left( {k,z - 1} \right)}} \right\|}^2}} \right\} \\ 
 \end{array}
\end{equation}
The objective function in (\ref{phae}) is obtained from the Q function in (\ref{finalqf}) ${\varepsilon _u} = \widehat\varepsilon _u^{\left( {k,z} \right)}$ with ${{\bf{h}}_u} = \widehat{\bf{h}}_u^{\left( {k,z - 1} \right)}$. 

Since we assume that the phase drifts $\left\{ {{\theta _{u,m}}} \right\}_{m = 1}^M$ are independent for different blocks, we can decouple the problem (\ref{phae}) into $M$ sub-problems
\begin{equation}\label{phae2}
\begin{array}{l}
\widehat\theta _{u,m}^{\left( {k,z} \right)} = \arg \mathop {\max }\limits_{{\theta _{u,m}}} \left\{ \begin{array}{l}
\\ 
\\ 
\end{array} \right. \\ 
\left. { - {{\left\| {\widehat{\bf{r}}_{u,m}^{\left( k \right)} - {e^{j{\theta _{u,m}}}}{\bf{\Gamma }}\left( {\widehat\varepsilon _u^{\left( {k,z} \right)}} \right){{\bf{F}}^H}{\bf{D}}\left( {{{\bf{m}}_{{{\bf{X}}_{u,m}}}}} \right){\bf{F}}\widehat{\bf{h}}_u^{\left( {k,z - 1} \right)}} \right\|}^2}} \right\} \\ 
\end{array}
\end{equation}
for $m = 1,2, \cdots ,M$, one for each block. Directly solving (\ref{phae2}) gives the phase estimate for the $m^{th}$ block:
\begin{equation}\label{phaest}
	\widehat\theta _{u,m}^{\left( {k,z} \right)} = \angle {\left[ {{\bf{\Gamma }}\left( {\widehat\varepsilon _u^{\left( {k,z} \right)}} \right){{\bf{F}}^H}{\bf{D}}\left( {{{\bf{m}}_{{{\bf{X}}_{u,m}}}}} \right){\bf{F}}\widehat{\bf{h}}_u^{\left( {k,z - 1} \right)}} \right]^H}\widehat{\bf{r}}_{u,m}^{\left( k \right)}
\end{equation}
With the new phase estimates $\left\{ {\widehat\theta _{u,m}^{\left( {k,z} \right)}} \right\}$, we go to the third stage.
\\

\noindent \emph{The Third Stage --- Channel Estimation}

\begin{figure*}[!t]
	\centering
	\includegraphics[width=6in]{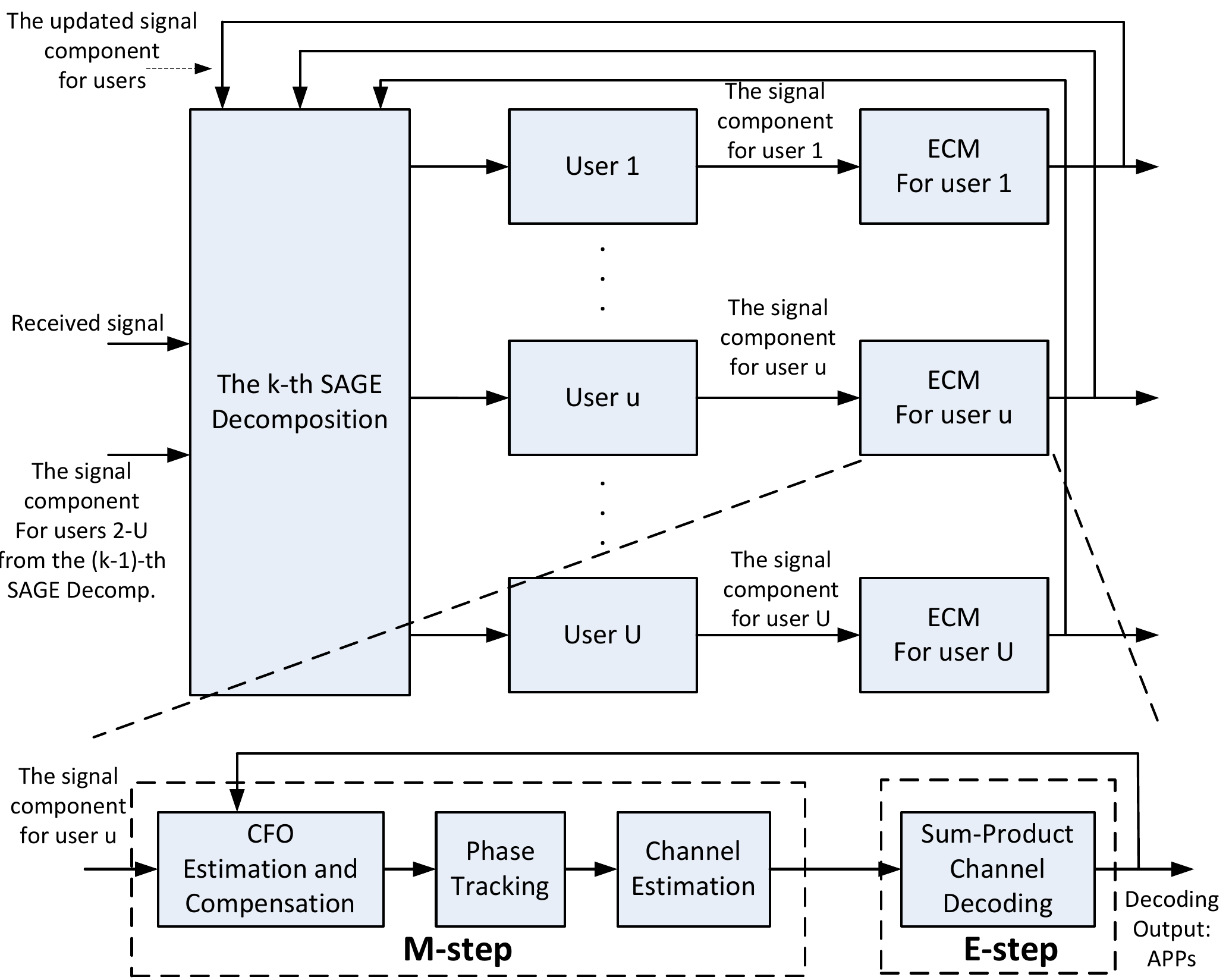}
	\caption{The operating flow chart for the proposed SAGE-ECM algorithm.} \label{sage_ecm_ill}
\end{figure*}  

The third state updates the channel gains with the CFO and the phase drifts fixed to their last estimates. The new set of parameter estimates after the third stage is $\widehat\Omega _u^{\left( {k,z} \right)} = \left\{ {\widehat\varepsilon _u^{\left( {k,z} \right)},\left\{ {\widehat\theta _{u,m}^{\left( {k,z} \right)}} \right\}_{m = 1}^M,\widehat{\bf{h}}_u^{\left( {k,z} \right)}} \right\}$, where the new channel estimates are given by 
\begin{equation}\label{chae}
\begin{array}{l}
\widehat{\bf{h}}_u^{\left( {k,z} \right)} = \arg \mathop {\max }\limits_{{{\bf{h}}_u}} \left\{ \begin{array}{l}
\\ 
\\ 
\end{array} \right. \\ 
\left. { - \sum\limits_{m = 1}^M {{{\left\| {\widehat{\bf{r}}_{u,m}^{\left( k \right)} - {e^{j\widehat\theta _{u,m}^{\left( {k,z} \right)}}}{\bf{\Gamma }}\left( {\widehat\varepsilon _u^{\left( {k,z} \right)}} \right){{\bf{F}}^H}{\bf{D}}\left( {{{\bf{m}}_{{{\bf{X}}_{u,m}}}}} \right){\bf{F}}{{\bf{h}}_u}} \right\|}^2}} } \right\} \\ 
\end{array}
\end{equation}
The objective function in (\ref{chae}) is obtained from the Q function in (\ref{finalqf}) with ${\varepsilon _u} = \widehat\varepsilon _u^{\left( {k,z} \right)}$, $\left\{ {{\theta _{u,m}} = \widehat\theta _{u,m}^{\left( {k,z} \right)}} \right\}_{m = 1}^M$. 

The solution for (\ref{chae}) is given by the least square (LS) estimate \cite{kay1998fundamentals}:
\begin{equation}\label{chaest}
\begin{array}{l}
\widehat{\bf{h}}_u^{\left( {k,z} \right)} \\ 
= \frac{1}{M}\sum\limits_{m = 1}^M {{e^{ - j{{\widehat\theta }_{u,m}}}}{{\bf{F}}^H}{{\bf{D}}^{ - 1}}\left( {{{\bf{m}}_{{{\bf{X}}_{u,m}}}}} \right){\bf{F}}{{\bf{\Gamma }}_m}\left( { - \widehat\varepsilon _u^{\left( {k,z} \right)}} \right)\widehat{\bf{r}}_{u,m}^{\left( k \right)}}  \\ 
\end{array}
\end{equation}
This finishes the final stage of the M-step. We then iterate back to the E-step with the set of new parameter estimates $\widehat\Omega _u^{\left( {k,z} \right)} = \left\{ {\widehat\varepsilon _u^{\left( {k,z} \right)},\left\{ {\widehat\theta _{u,m}^{\left( {k,z} \right)}} \right\}_{m = 1}^M,\widehat{\bf{h}}_u^{\left( {k,z} \right)}} \right\}$.
\\

\noindent \textbf{Initialization and Termination of ECM Iteration}:

We bootstrap the ECM iteration with initial estimates $\widehat\Omega _u^{\left( {k,0} \right)} = \left\{ {\widehat\varepsilon _u^{\left( {k,0} \right)},\left\{ {\widehat\theta _{u,m}^{\left( {k,0} \right)}} \right\}_{m = 1}^M,\widehat{\bf{h}}_u^{\left( {k,0} \right)}} \right\}$, where the initial CFO estimate and channel estimates $\widehat\varepsilon _u^{\left( {k,0} \right)}$, $\widehat{\bf{h}}_u^{\left( {k,0} \right)}$ are obtained from the preamble of user $u$; the initial phase estimate $\widehat\theta _{u,m}^{\left( {k,0} \right)}$ is obtained from the pilot subcarriers of the $m^{th}$  block for every $m$.

We repeat the E-step and M-step of ECM iteratively. When the number of iterations $z$ reaches the preset maximum limit $Z$, we terminate the ECM algorithm for user $u$ and take the final estimates as the approximate solution for the joint estimation problem (8):
$$\begin{array}{l}
\left\{ {\hat \varepsilon _u^{\left( k \right)},\left\{ {\hat \theta _{u,m}^{\left( k \right)}} \right\}_{m = 1}^M,\widehat{\bf{h}}_u^{\left( k \right)},\widehat{\bf{X}}_u^{\left( k \right)}} \right\} \\ 
= \left\{ {\hat \varepsilon _u^{\left( {k,Z} \right)},\left\{ {\hat \theta _{u,m}^{\left( {k,Z} \right)}} \right\}_{m = 1}^M,\widehat{\bf{h}}_u^{\left( {k,Z} \right)},\widehat{\bf{X}}_u^{\left( {k,Z} \right)}} \right\} \\ 
\end{array}$$
where  $\left\{ {\widehat\varepsilon _u^{\left( {k,Z} \right)},\left\{ {\widehat\theta _{u,m}^{\left( {k,Z} \right)}} \right\}_{m = 1}^M,\widehat{\bf{h}}_u^{\left( {k,Z} \right)}} \right\}$ are the final parameter estimates obtained from the M-step of the $Z^{th}$ ECM iteration,   $\widehat{\bf{X}}_u^{\left( {k,Z} \right)}$ are obtained by making hard decisions based on the symbol-wise APPs $\left\{ {p\left( {{X_{u,m,i}}\left| {\widehat{\bf{r}}_u^{\left( k \right)},\widehat\Omega _u^{\left( {k,Z} \right)},{C_u}} \right.} \right)} \right\}$. A operating flow chart for the proposed SAGE-ECM algorithm is shown in Fig. \ref{sage_ecm_ill}.

\subsection{Alternative Iterative Receiver}

We have described our proposed SAGE-ECM receiver in Section III.C.  We can obtain an alternative receiver by adopting another message passing algorithm --- the min-sum algorithm --- for channel decoding.

Let us revisit (\ref{subprob}) and consider how to solve it iteratively. We can apply SAGE again in this single-user subproblem. Specifically, we can break up one iteration into several stages where one stage updates just one parameter while fixing all other parameters to their last estimates. This iterative receiver corresponds to a pure SAGE framework, where all variables are treated as parameters and updated in a sequential manner.

Using the above idea, the $z^{th}$ iteration for solving (\ref{subprob}) consists of four stages. In the first, second and third stages, the CFOs, the phases and the channel gains are updated to $\widehat\varepsilon _u^{\left( {k,z} \right)}$,  $\left\{ {\theta _{u,m}^{\left( {k,z} \right)}} \right\}_{m = 1}^M$ and ${\widehat{\bf{h}}_u^{\left( {k,z} \right)}}$  employing the methods of (\ref{cfoest}), (\ref{phaest}) and (\ref{chaest}), respectively. The fourth stage updates the transmit symbols by 
\begin{equation}\label{ecm23}
\begin{array}{l}
\widehat{\bf{X}}_u^{\left( {k,z} \right)} = \arg \mathop {\max }\limits_{{{\bf{X}}_u}} \left\{ \begin{array}{l}
\\ 
\\ 
\end{array} \right. \\ 
\left. { - \sum\limits_{m = 1}^M {{{\left\| {\widehat{\bf{R}}_{u,m}^{\left( k \right)} - {e^{j\widehat\theta _{u,m}^{\left( {k,z - 1} \right)}}}{\bf{D}}\left( {{{\bf{X}}_{u,m}}} \right){\bf{F}}\widehat{\bf{h}}_u^{\left( {k,z} \right)}\left( {{{\widehat\Omega }_u^{\prime  \left( {k,z - 1} \right)}}} \right)} \right\|}^2}} } \right\} \\ 
\end{array}
\end{equation}
where $\widehat{\bf{R}}_{u,m}^{\left( k \right)} = {\bf{F}}\left( {{\bf{\Gamma }}\left( { - \widehat\varepsilon _u^{\left( {k,z - 1} \right)}} \right)\widehat{\bf{r}}_{u,m}^{\left( k \right)}} \right)$ is the vector of the frequency-domain signals after CFO compensation. With $\left\{ {\hat \varepsilon _u^{\left( {k,z} \right)},\left\{ {\hat \theta _{u,m}^{\left( {k,z} \right)}} \right\}_{m = 1}^M,{\widehat{\bf{h}}_u^{\left( {k,z} \right)}},\widehat{\bf{X}}_u^{\left( {k,z} \right)},} \right\}$, we complete the $z^{th}$ iteration.  

The fourth stage in (\ref{ecm23}) corresponds to the channel decoding operation. In contrast to the channel decoding in the SAGE-ECM framework in Section III.C, where the sum-product algorithm was used to obtain the APPs of the transmit symbols, the  channel decoding method corresponding to (\ref{ecm23}) is the min-sum algorithm \cite{wiberg1996codes, kschischang2001factor}. For example, if convolution codes are employed, the SAGE-ECM framework (in Section III.C)  would use the BCJR algorithm, but the pure SAGE framework  here would use the Viterbi algorithm (a special case of the min-sum algorithm \cite{wiberg1996codes, kschischang2001factor}).

Henceforth, we will refer to the iterative receiver in Section III.C as the SAGE-ECM Sum-Product Rx, and the iterative receiver here as the SAGE Min-Sum Rx. We compare their performances through simulation study in the next section.   

The key difference between the SAGE Min-Sum Rx in this section and the SAGE-ECM Sum-Product Rx in Section III.C is how channel decoding assists the channel-parameter estimation. For the SAGE Min-Sum Rx, the symbol estimates $\widehat{\bf{X}}_u^{\left( {k,z} \right)}$ given by min-sum channel decoding are hard decisions --- we lose soft information on ${{\bf{X}}_u}$ that specifies the levels of confidence on our estimates. As will be seen in  the next section, these hard decisions cause significant performance degradation when accepted as parameter estimates. The SAGE-ECM Sum-Product Rx assists the channel-parameter estimation using the soft information on the transmit symbols in a more sophisticated manner. Finally, we remark that the SAGE min-sum RX here is equivalent to the receiver proposed in \cite{pun2007iterative} (see a detail discussion about \cite{pun2007iterative} in Appendix C).

\subsection{Complexity Comparison}

We analyze and compare the complexities of our SAGE-ECM receiver and other receivers. For these receivers, which are OFDM based, the most computation intensive parts are the DFT/IDFT operations, expressed by the multiplications with matrix ${\bf{F}}$ or ${{\bf{F}}^H}$ in the equations of this paper. For implementation, the DFT/IDFT operations can be realized using FFT/IFFT. The complexity of FFT/IFFT is $O\left( {N\log N} \right)$. The other operations involved in the receivers are scalar additions, multiplications, divisions whose complexities are linear with the frame size. Therefore, for comparison purposes, we count the numbers of FFT and IFFT operations in our SAGE-ECM receiver and other receivers.     

We first look at a traditional non-iterative estimation receiver where the channel gains and CFOs are estimated from the preamble, phases are estimated from the pilot subcarriers, and then an iterative interference cancellation approach (such as the one proposed in \cite{dang2012experimental}) is used for decoding. There are $2U$ preamble symbols and $M' = M - 2U$ data symbols in each frame. To process the preambles of all users, it is required to perform $2U$ FTT operations. Then, in each iteration of iterative interference cancellation, the decoding of one user needs to perform $M'$ FFT for transforming the interference-cancelled time-domian signals to the frequency domain for decoding, and perform $M'$ IFFT for transforming the decoded signal back to the time domain to be used to cancel interference by the next user. The number of iterations for interference cancellation is $K$. Totally, there are $2U + 2KUM'$ FTT/IFFT operations for this traditional non-iterative estimation receiver. 

We next look at our SAGE-ECM receiver. For the ECM iteration of one user, we need to perform channel decoding in the frequency domain; then convert the $M'$  mean vectors of frequency-domain data symbol vectors to the time domain for parameter estimations; after that,  we convert the $M'$ compensated time-domain signals back to the frequency domain for  channel decoding in the next round of ECM iteration. Therefore, within one ECM iteration of one user, we need to perform  $2M'$ FFT/IFFT operations. Since we have $U$ users, $Z$ ECM iterations and $K$ SAGE iterations, in total we need $2KZUM'$ FTT/IFFT operations for all the SAGE-ECM iterations. To initialize the SAGE-ECM iteration, we also need to process the preambles of users. This also needs $2U$ FTT operations. Totally, there are  $2U + 2KZUM'$ FTT/IFFT operations for the SAGE-ECM receiver. From these results, we can conclude that the computation complexity of our SAGE-ECM receiver is approximately  $Z$ times larger than that of the traditional non-iterative estimation receiver. 

The pure SAGE receiver discussed in Section III.D needs the same number of FFT/IFFT operations as our SAGE-ECM receiver. Compared with our SAGE-ECM receiver, the complexity of the pure SAGE receiver is just slightly less computation intensive thanks to its simpler  min-sum channel decoding: both receivers, however, have the same order of complexity in the frame size.

\section{Simulation and Experimental Results}

This section presents simulation results and experimental results on software defined radio.

\subsection{Simulation Results}

In all simulations, the frame format is a slightly modified version of the 802.11a frame format \cite{ieee2007wlan}. The DFT size is $N = 64$. The CP is of length ${N_{cp}} = 16$. Among the $N = 64$ subcarriers, there are ${N_d} = 48$ data subcarriers, ${N_p} = 2U$ ($U$ is the number of users) pilot subcarriers and $64-N_d-N_p$ unused guard band subcarriers. Each user transmits known symbols on 2 of the $2U$ pilot subcarriers, and nulls the signal on the other $2U-2$ pilot subcarriers. The modulation is BPSK. A way to realize a low-rate channel code scheme for IDMA is to serially concatenate a forward error correction (FEC) code with a repetition code \cite{ping2005interleave}.  In our simulation, the FEC is a regular Repeat Accumulate (RA) code \cite{divsalar1998coding} with code rate ${R_1} = {1 \mathord{\left/
		{\vphantom {1 3}} \right.
		\kern-\nulldelimiterspace} 3}$, and the code rate of the repetition code is ${R_2} = {1 \mathord{\left/
		{\vphantom {1 U}} \right.
		\kern-\nulldelimiterspace} U}$. Therefore, the overall code rate is $R = {R_1}{R_2} = {1 \mathord{\left/
		{\vphantom {1 9}} \right.
		\kern-\nulldelimiterspace} 3U}$. The interleavers of all three users are generated randomly. The payload of each frame has 2400 bits. The two preamble blocks for each user are two successive copies of the long training sequence (LTS) defined in the 802.11a standard. However, as described in Section II.A, the two preambles of different users occupy different block times.

We set the length of the discrete channel vectors for all three users to ${L_1} = {L_2} = {L_3} = 4$. The channel taps are generated according to the channel mode given in Section II.B. We assume that the receiver of the base station can capture the first channel path of user 1, so that the timing mismatch between user 1 and the base station is set to ${\mu _1} = 0$. Furthermore, other ${\mu _u}$ are randomly chosen from the interval $\left[ {0,9} \right]$. Thus, the loose time synchronization requirement is satisfied in simulations. We set the CFO of each user to $\rho $ or $-\rho $ with equal probability, where $\rho$ is the so-called CFO attenuation factor \cite{huang2005interference},  and it is a deterministic parameter ranging in $\left[ {0,0.5} \right]$. In our simulations, we vary the value of $\rho$ to investigate its impact on system performance. SNR is defined as ${{{E_b}} \mathord{\left/
		{\vphantom {{{E_b}} {{N_0}}}} \right.
		\kern-\nulldelimiterspace} {{N_0}}}$, where ${E_b}$ is the energy per information bit and ${N_0}$ is the noise variance. All simulation results presented here are obtained by averaging over ${\rm{3}} \times {\rm{1}}{{\rm{0}}^{\rm{3}}}$ frames.

For performance comparison, we investigate the following four approaches for OFDM-IDMA systems: (i) the iterative interference cancellation approach in \cite{dang2012experimental} with perfect knowledge of channel parameters (Full CSI Rx); (ii) the iterative interference cancellation approach in \cite{dang2012experimental} with the one-shot channel parameter estimations obtained from the preambles and pilots (One-shot Est. Rx); (iii) the SAGE Min-Sum Rx approach described in Section III.D as a benchmark; and (iv) the SAGE-ECM Sum-Product Rx approach proposed in Section III.C. For all receivers, we performed iterations until the algorithm converged (with SAGE iteration number $K=10$, and ECM iteration number $Z=20$). We remark that the performance of One-shot Est. Rx is the initial point of SAGE-ECM Sum-Product Rx.

\begin{figure*}
	\begin{minipage}[t]{0.5\linewidth}
		\centering
		\includegraphics[width=1\textwidth]{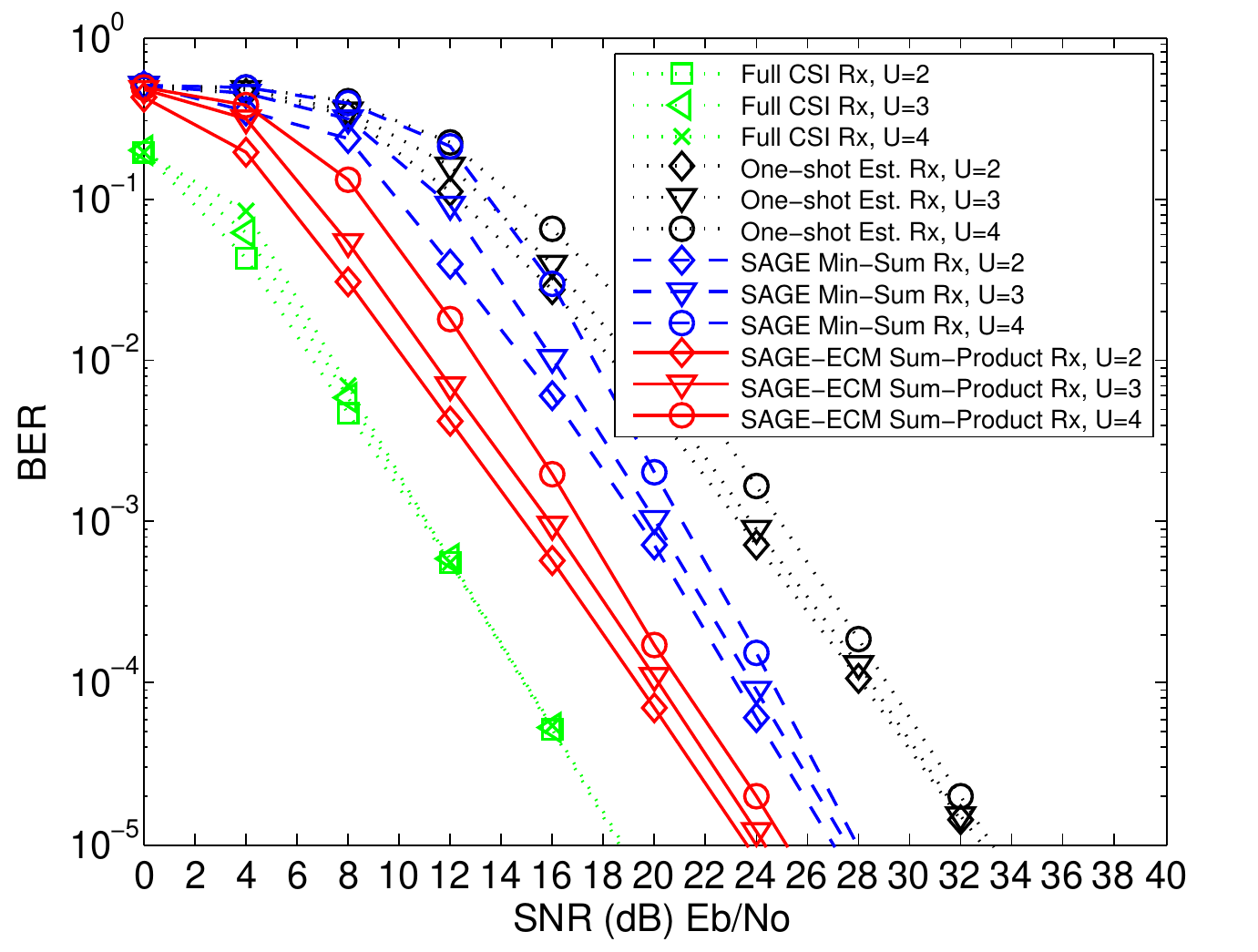}
		\caption{BER versus SNRs with $\rho=0.2$.}
		\label{cha3simu1}
	\end{minipage}
	\begin{minipage}[t]{0.5\linewidth}
		\centering
		\includegraphics[width=1\textwidth]{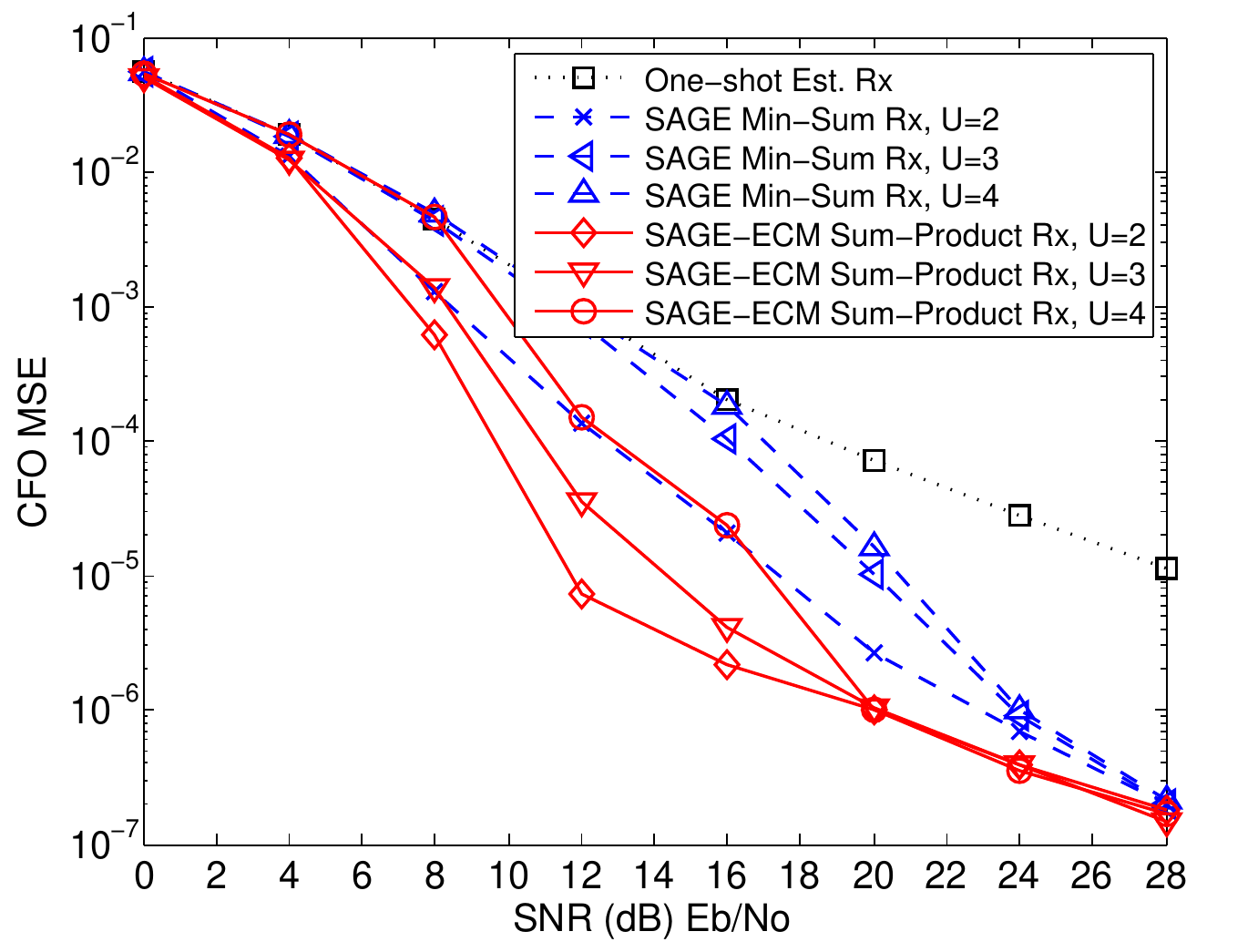}
		\caption{CFO MSEs versus SNRs with  $\rho=0.2$.}
		\label{cha3simu2}
	\end{minipage}
\end{figure*}

Fig. \ref{cha3simu1} presents the results of bit error rate (BER) versus SNR. The CFO attenuation factor is fixed to $\rho  = 0.2$. The systmes with users $U=2,3,4$ are simulated. From the results, we first note that the BER gap between Full CSI Rx and One-shot Est. Rx is large. Specifically, the estimation errors of one-shot estimation induce around 13 dB SNR loss at  BER = ${10^{ - 5}}$. Intuitively, since each imperfect parameter estimate induces an SNR penalty and we have many channel parameters in the system, the SNR penalties accumulate to an overall large penalty. 

The joint channel-parameter estimation, CFO compensation and channel decoding approachs reduce the performance loss, as shown in Fig. \ref{cha3simu1}. Compared to one-shot Est. Rx, SAGE-ECM Sum-Product Rx (proposed) yields 8dB SNR improvement and SAGE Min-Sum Rx (benchmark) yields 5dB SNR improvement at BER = ${10^{ - 5}}$. The 3 dB performance gap between SAGE-ECM Sum-Product Rx and SAGE Min-Sum Rx justifies that the belief about data symbol given by the sum-product channel decoding can assist the channel-parameter estimation procedures in a better manner.

\begin{figure*}
	\begin{minipage}[t]{0.5\linewidth}
		\centering
		\includegraphics[width=1\textwidth]{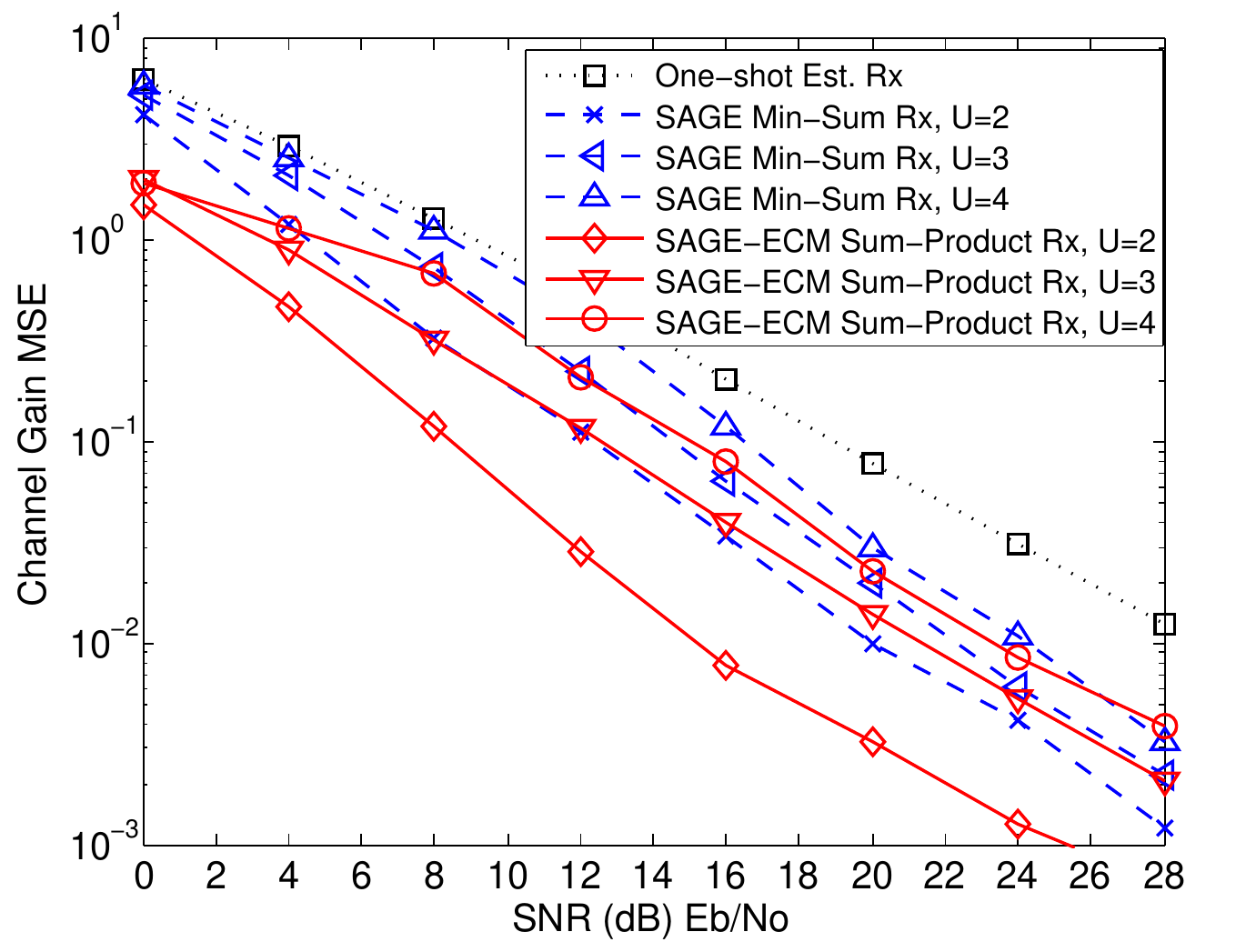}
		\caption{Channel gain MSEs versus SNRs with $\rho=0.2$. }
		\label{cha3simu3}
	\end{minipage}	
	\begin{minipage}[t]{0.5\linewidth}
		\centering
		\includegraphics[width=1\textwidth]{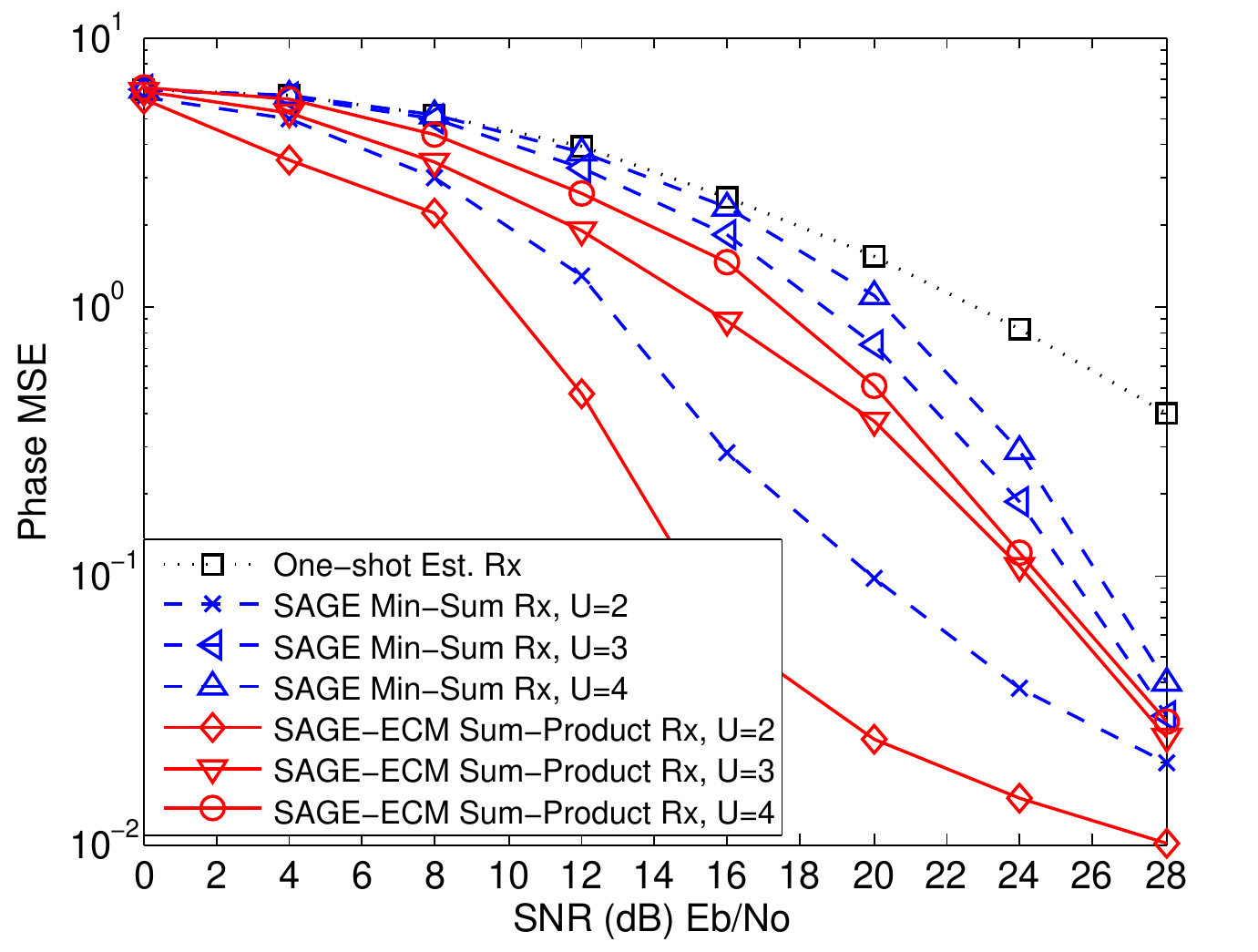}
		\caption{Phase MSEs versus SNRs with  $\rho=0.2$.}
		\label{cha3simu4}
	\end{minipage}
\end{figure*}

We also evaluate the mean square error (MSE) of the estimated channel parameters.  Fig. \ref{cha3simu2}, Fig. \ref{cha3simu3} and Fig. \ref{cha3simu4} present the MSEs of the estimated CFOs, channels and phases versus SNR. The CFO attenuation factor is fixed to $\rho  = 0.2$. The units of the CFOs and the phases are Hz and radian. The systmes with users $U=2,3,4$ are simulated. From the MSE results Fig. \ref{cha3simu2}-\ref{cha3simu4}, we clearly see that the SAGE-ECM approaches do have more accurate estimates than the traditional preamble/pilot-based one-shot estimations. Moreover, the estimations of SAGE-ECM Sum-Product Rx are better than that of SAGE Min-Sum Rx; the difference in MSEs between them decreases as the SNR increases. The reason is that the results of min-sum decoding are hard decisions. In the low SNR regime, the hard decisions are not reliable enough and will propagate the decoding errors to channel parameter estimations. In the high SNR regime, the decoding results of min-sum decoding approaches that of sum-product decoding, thus the error gaps in MSEs become narrow. 


\begin{figure*}
	\begin{minipage}[t]{0.5\linewidth}
		\centering
		\includegraphics[width=1\textwidth]{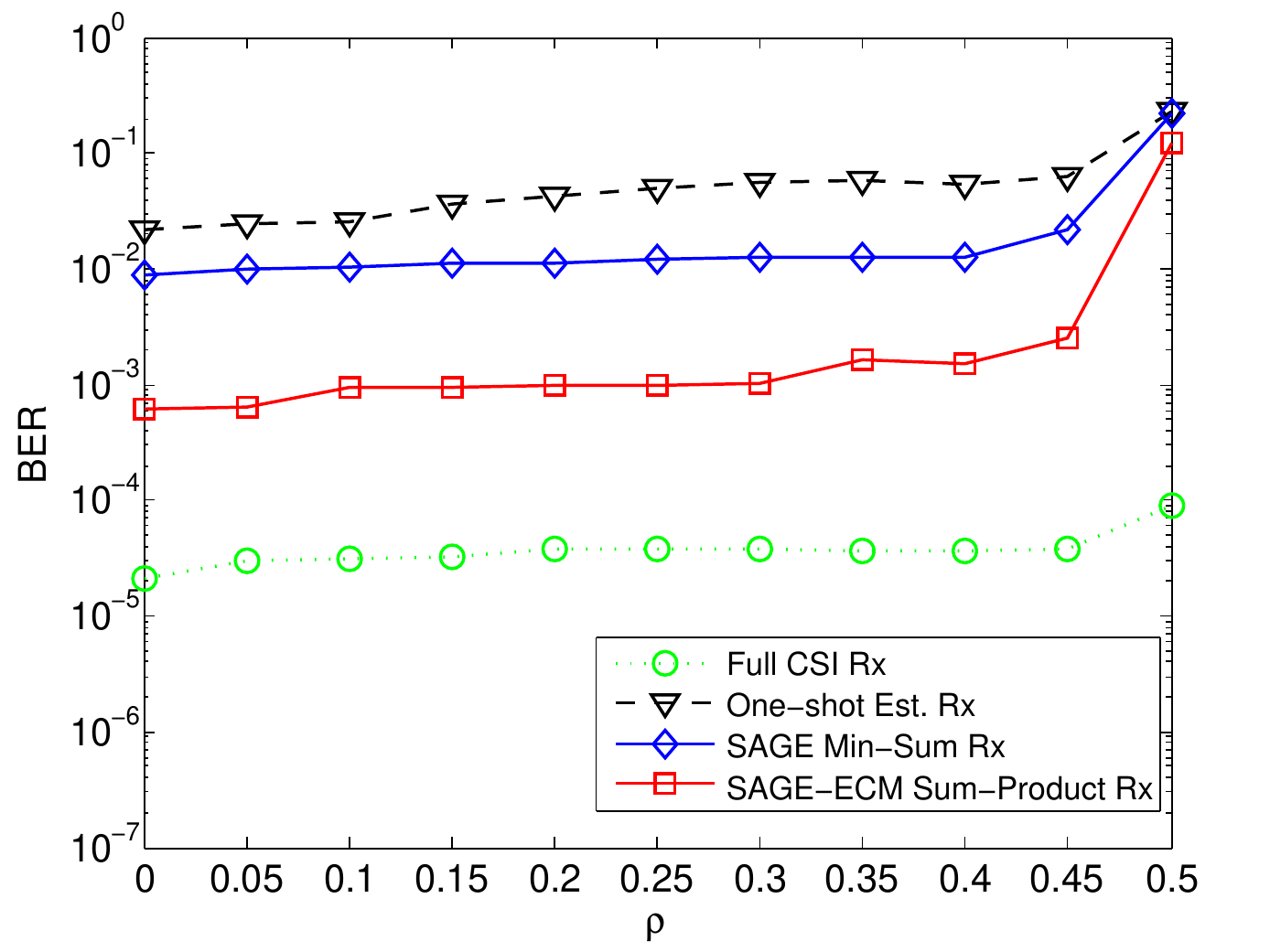}
		\caption{BER versus $\rho$ with  SNR=16 dB, $U=3$.}
		\label{cha3simu5}
	\end{minipage}
	\begin{minipage}[t]{0.5\linewidth}
		\centering
		\includegraphics[width=1\textwidth]{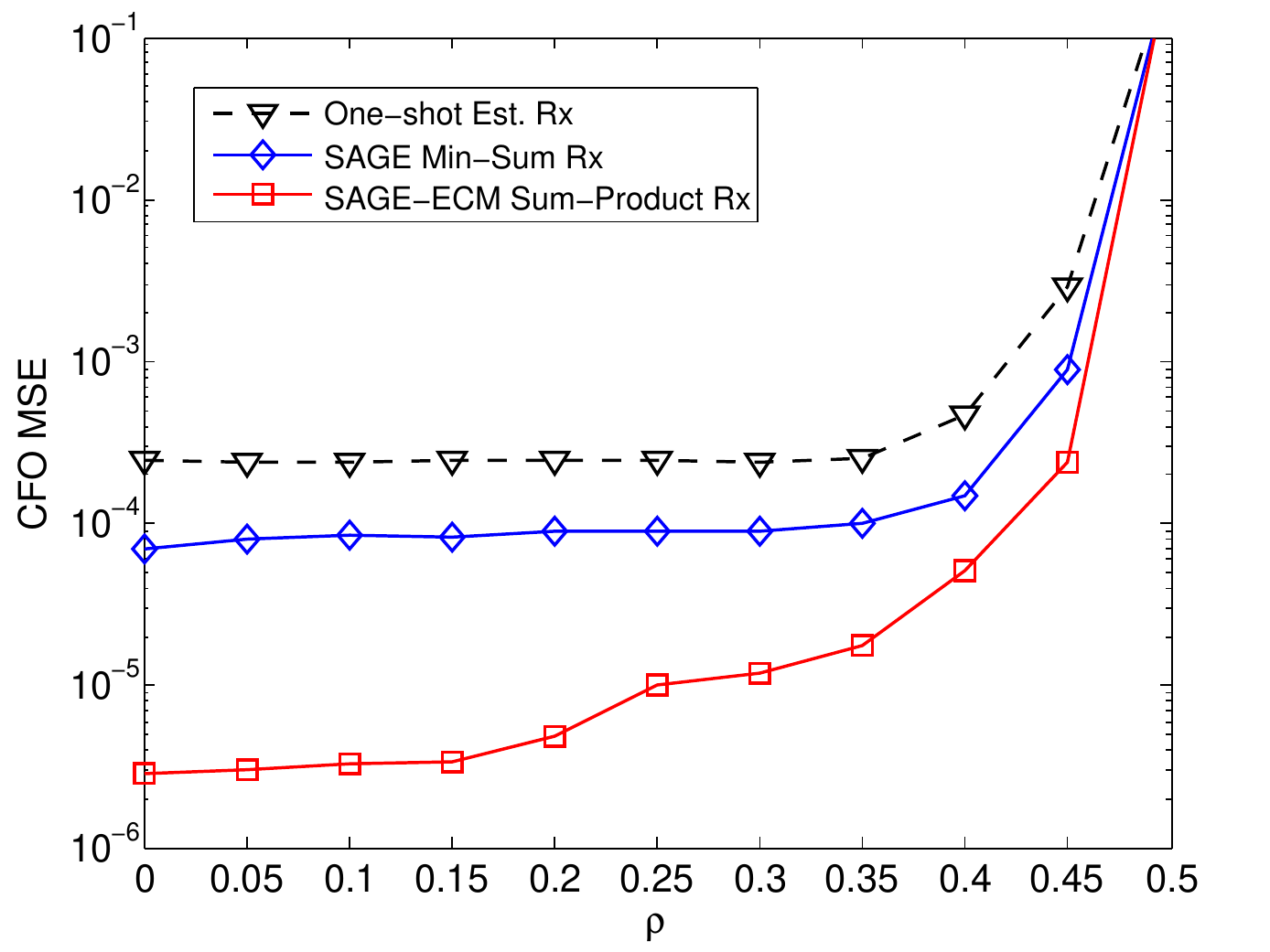}
		\caption{CFO MSEs versus $\rho$ with SNR=16 dB, $U=3$.}
		\label{cha3simu6}
	\end{minipage}
\end{figure*}

Fig. \ref{cha3simu5} shows the impact of CFO attenuation factor $\rho$ on BER, where SNR is fixed to 16 dB and the system includes $U=3$ users. The results again confirm the better performance of SAGE-ECM Sum-Product Rx over One-shot Est. Rx and SAGE Min-Sum Rx. It also shows that the BERs of all approaches are insensitive to $\rho$: note that the different approaches have different performances; just that the performance of each approach is not sensitive to $\rho$. This result implies that it is not the magnitude of  that affects performance; it is the estimation errors of $\rho$, which vary among the different approaches.

\begin{figure*}
	\begin{minipage}[t]{0.5\linewidth}
		\centering
		\includegraphics[width=1\textwidth]{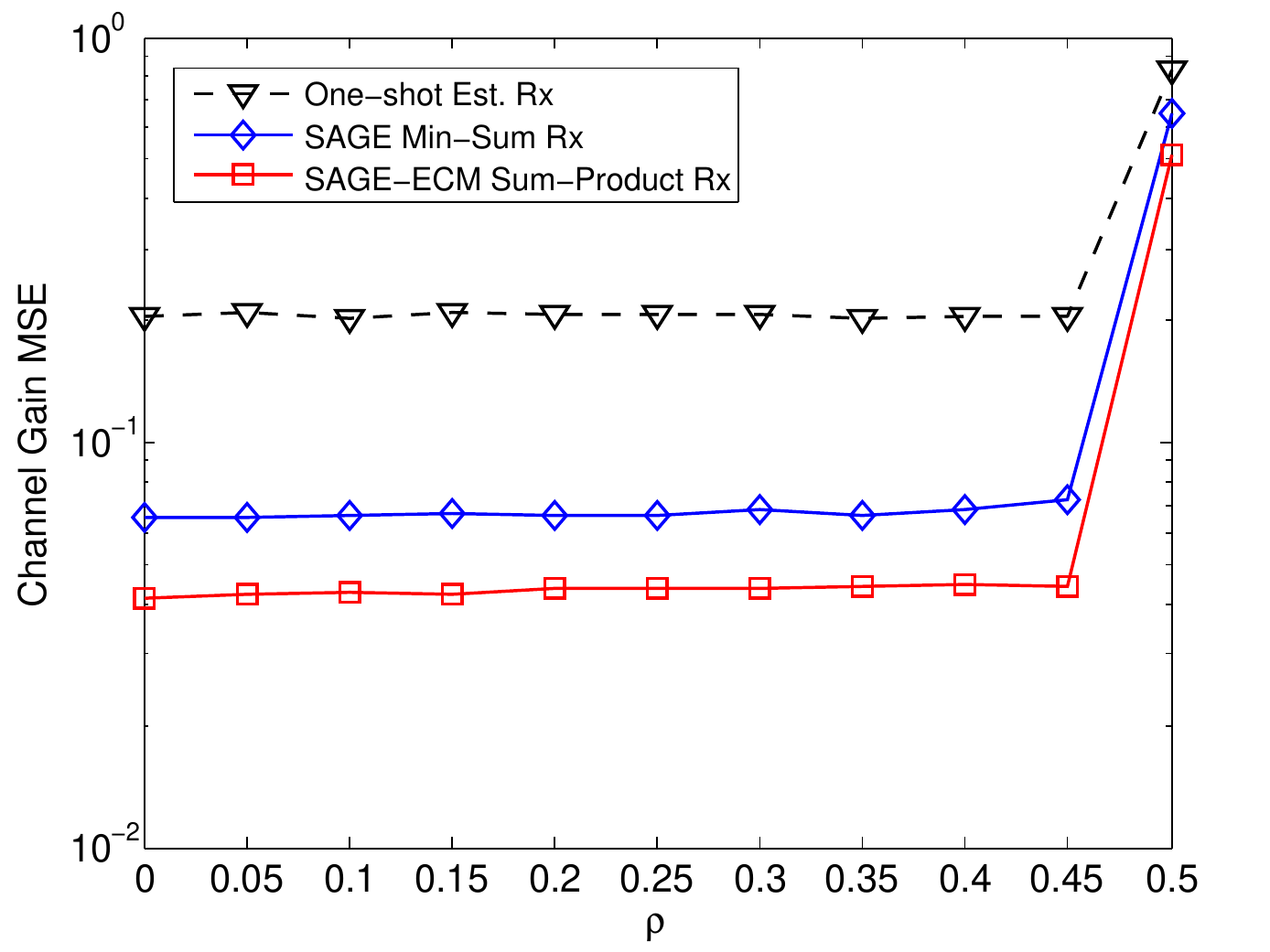}
		\caption{Channel gain MSEs versus $\rho$ with SNR=16 dB, $U=3$.}
		\label{cha3simu7}
	\end{minipage}
	\begin{minipage}[t]{0.5\linewidth}
		\centering
		\includegraphics[width=1\textwidth]{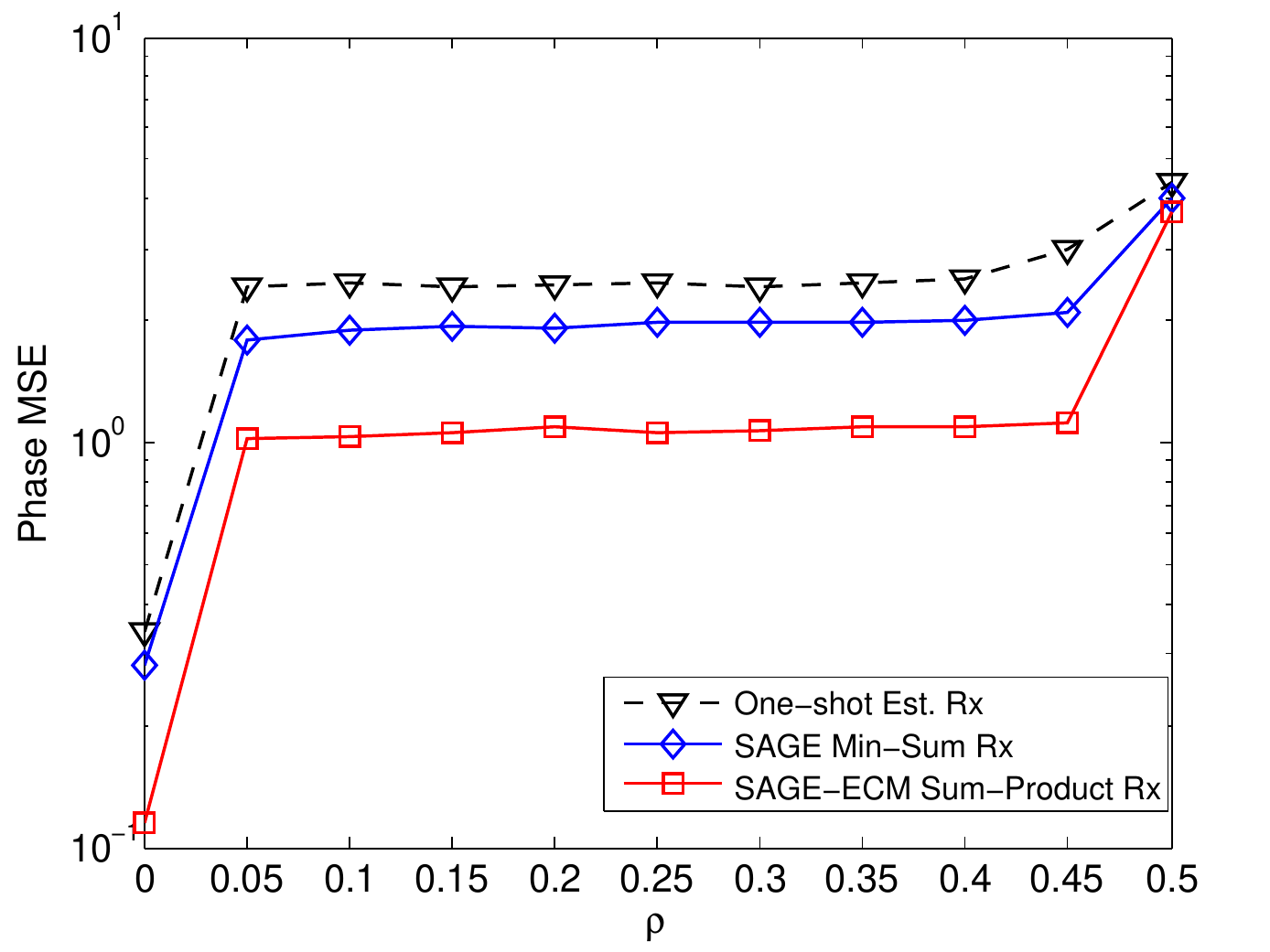}
		\caption{Phase MSEs versus $\rho$ with SNR=16 dB, $U=3$.}
		\label{cha3simu8}
	\end{minipage}
\end{figure*}

\begin{figure*}
	\begin{minipage}[t]{0.5\linewidth}
		\centering
		\includegraphics[width=3.5in]{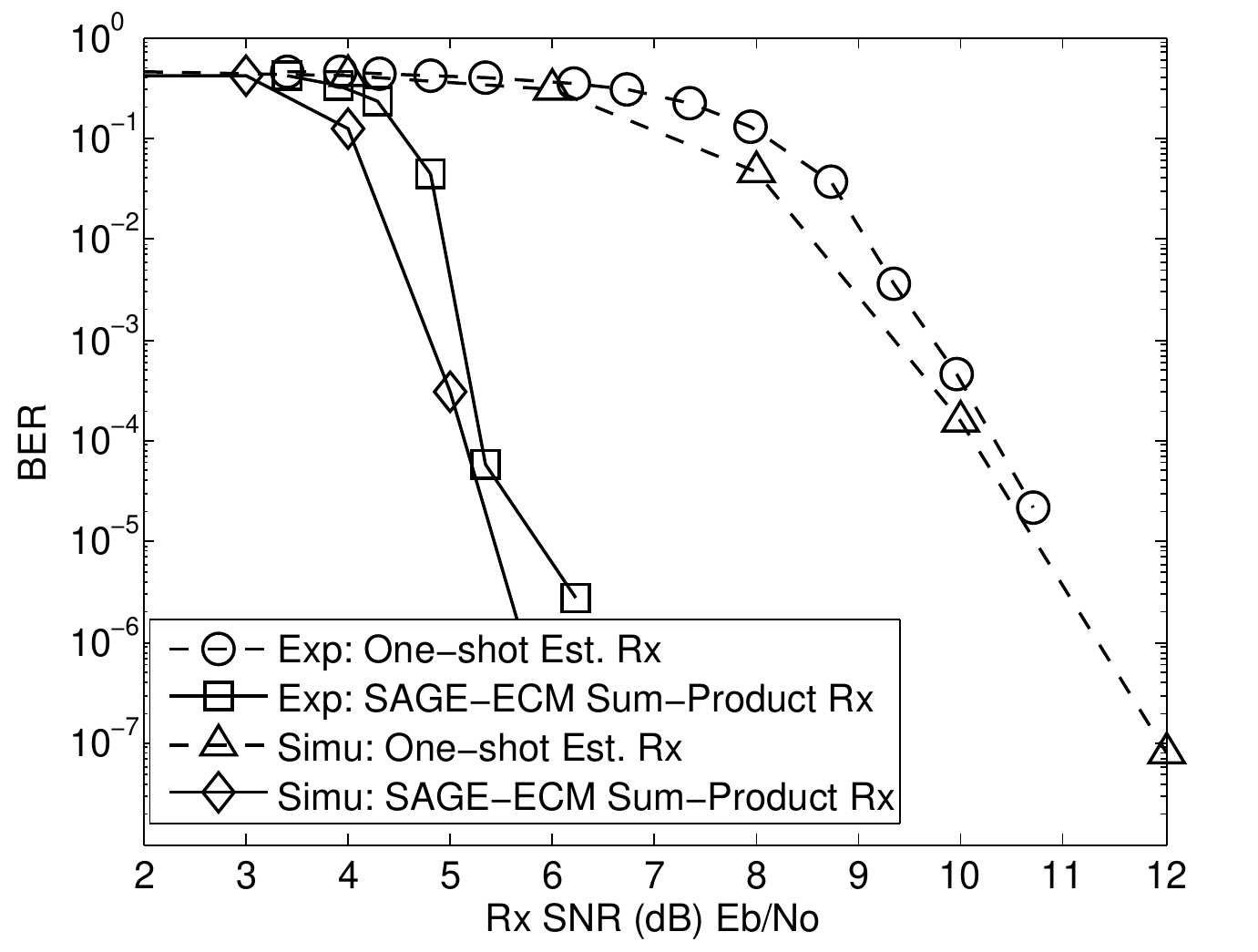}
		\caption{Experimental BER results with $U=2$.}
		\label{cha3simu9}
	\end{minipage}
	\begin{minipage}[t]{0.5\linewidth}
		\centering
		\includegraphics[width=3.5in]{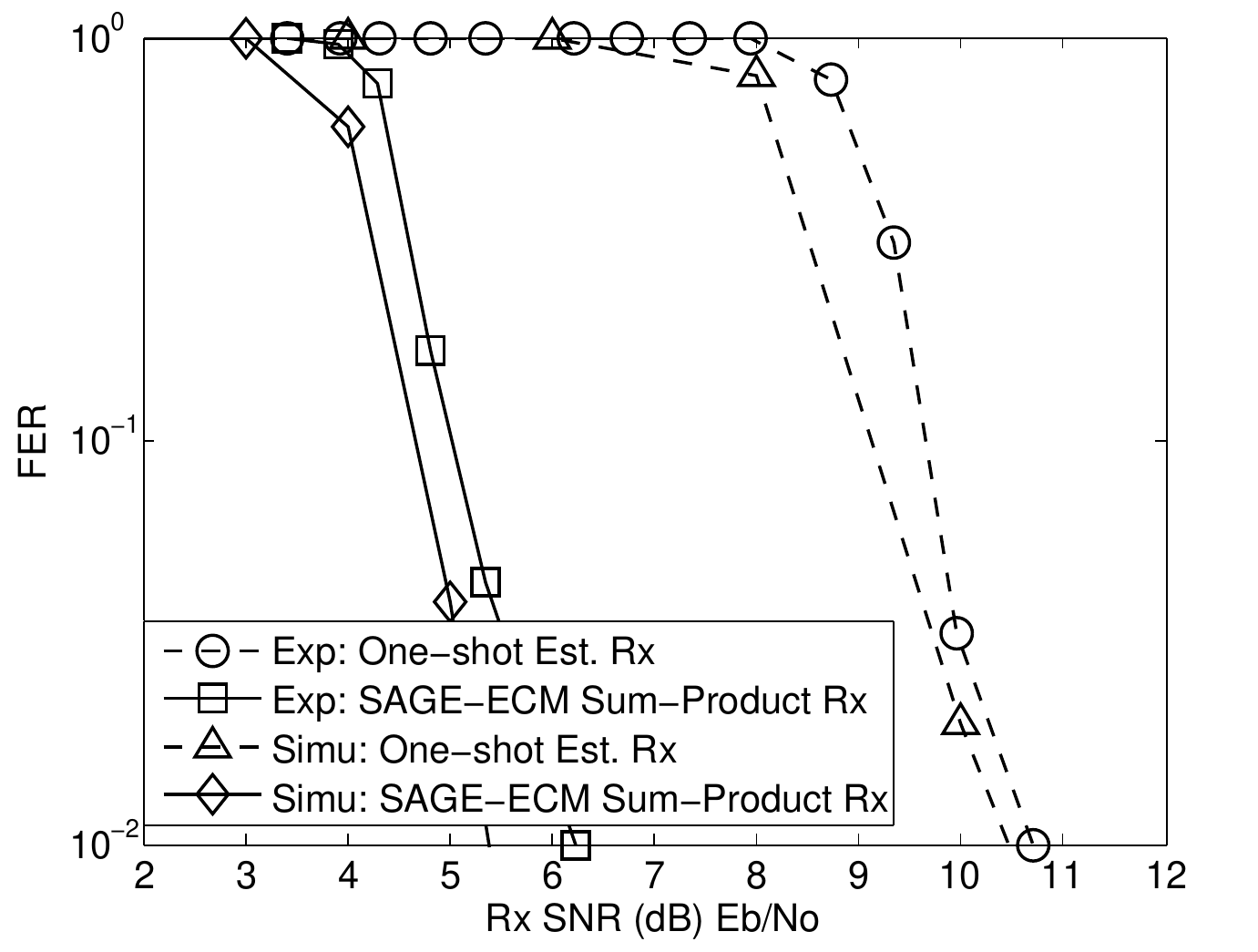}	
		\caption{Experimental FER results with $U=2$.}
		\label{cha3simu10}
	\end{minipage}
\end{figure*}

Fig. \ref{cha3simu6}, Fig. \ref{cha3simu7} and Fig. \ref{cha3simu8}  show the MSEs of the estimated CFOs, channels and phases versus  $\rho$. The SNR is fixed to 16 dB. The system includes $3$ users. The performance trends are the same with the BER in terms of $\rho$, as shown in Fig. \ref{cha3simu5}. SAGE-ECM Sum-Product Rx has better MSE performances than SAGE Min-Sum Rx and One-shot Est. Rx do. For all approaches, their estimation errors depends only weakly on $\rho$ for a wide range of $\rho$; performances only degrade as $\rho$ is near 0.5 (the worst case). We thus conclude that it is the error in the estimation of CFO that has more effects than the actual CFO value itself.

\subsection{Experimental Results}

Going beyond simulations, we also verify our proposed approach experimentally. We implemented an OFDM-IDMA system using a software defined radio (SDR) platform. We collected the data for the received signal from the experimental system and evaluate the performance of the proposed approach using the collected data.

The experimental system is built on the USRP N210 hardware \cite{usrp} and the GNU Radio software with the UHD hardware driver \cite{gnu}. We emulated an OFDM-IDMA system that includes one base station and two users ($U = 2$), by deploying three sets of USRP N210 with XCVR2450 boards \cite{usrp} in our lab.\footnote{We performed the experiment for OFDM-IDMA using the SDR prototype of the physical-layer network coding (PNC) systems reported in detail in \cite{lu2012implementation, lu2013real}. In the uplink of PNC, two users transmit signals to the relay simultaneously. This is similar to multiple-access systems. Therefore, we can borrow it for our use here.}  The base station used 802.11 channel 1 (2.412GHz) to poll the two users to transmit together at channel 11 (2.462GHz), thereby achieving the loose-time synchronization mentioned in Section II.C. The wireless channel bandwidth of our network is 4MHz. The use of 4 MHz bandwidth rather than a wider bandwidth is due to the limitation of the USRP hardware. After the terminal users received a beacon from the base station, they then transmit their signals simultaneously. The base station received the simultaneous transmissions and converted them to digital data for processing. We performed controlled experiments for different received SNRs. The receive powers of frames from two users at the base station were adjusted to be balanced (power imbalance within 1dB). The base station transmits 100 beacons to trigger 100 simultaneous uplink transmissions for each fixed SNR.

The frame format used is similar to the one used in simulations except for the following two differences. First, in the experimental setup, there are four pilot subcarriers within each OFDM block (as opposed to six in simulations) and each user transmits pilot symbols over two of them. Second, the channel coding scheme used is only the  ${1 \mathord{\left/
		{\vphantom {1 3}} \right.
		\kern-\nulldelimiterspace} 3}$
coding rate RA code without the repetition code. Each frame includes 256 OFDM blocks.  

The experimental BER results are shown in Fig. \ref{cha3simu9}. The frame error rate (FER) results are shown in Fig. \ref{cha3simu10}.  We compare the performances of SAGE-ECM Sum-Product Rx with One-shot Est. Rx. Generally, the performance trends are similar to those in our simulation results. Specifically, we observe that SAGE Sum-Product Rx achieves around 5 dB SNR gain over One-shot Est. Rx. at the BER of ${10^{ - 5}}$ in the experiment. We note that the shapes of the experimental BER curves are different from those of the simulations in Fig. \ref{cha3simu1}. The reason is that the channel in our experimental environment is rather flat over the 4 MHz bandwidth. To verify our experimental results, we perform an additional simulation where the system includes $U=2$ users, the frame format is the same as that in our experiment, the channel has one Rayleigh path (thus it is flat), and the CFOs are set to ${\left[ {{\varepsilon _1},{\varepsilon _2},} \right]^T} = {\left[ {{\rm{0}}{\rm{.06}},{\rm{0}}{\rm{.11}}} \right]^T}$ that are the means of the measured CFO values in our experiment.  The simulated BER and FER results under this setup are presented in Fig. \ref{cha3simu9} and Fig. \ref{cha3simu10}, respectively. As can be seen, the simulated results are consistent with the experimental results in that there is also around 5 dB SNR gain by SAGE-ECM Sum-Product Rx over One-shot Est. Rx.

\section{Conclusion}

We have put forth a framework that employs the SAGE and ECM algorithms to solve the problem of joint channel-parameter estimation, CFO compensation and channel decoding in multiuser OFDM-IDMA systems. Our framework is motivated by the fact that for multiuser OFDM systems, (i) one-shot non-iterative parameter estimation does not yield satisfactory accuracy; and (ii) one-shot non-iterative CFO compensation is impossible. For these reasons, we propose to solve the overall problem of channel-parameter estimation, CFO compensation, and channel decoding  jointly and iteratively.

Within our framework, we compared different schemes of assigning the role of hidden data in ECM, and concluded that treating the data symbols (as opposed to channel parameters) as the hidden data in ECM leads to better performance. For the consistency and completeness of this scheme, we bridge the time-domain channel-parameter estimation procedure and the frequency-domain channel decoding procedure using ``soft IDFT''. 

Our simulation results and real SDR experimental results indicate that compared with the traditional multiuser approach, the proposed approach can obtain 5 dB SNR gain for the two-user case, and 8 dB SNR gain for the three-user case.

\appendix

\subsection{Derivation of (\ref{sqf2})}

We first expand the sample-wise Q function in (\ref{sqf}) as 
\begin{equation}\label{A1}
\begin{array}{l}
 {Q_{m,i}}\left( {{\Omega _u}\left| {\hat \Omega _u^{\left( {k,z - 1} \right)}} \right.} \right) \\ 
 {\rm{ = }}\sum\limits_{{y_{u,m,i}}} {\log p\left( {\hat r_{u,m,i}^{\left( k \right)}\left| {{\Omega _u},{y_{u,m,i}}} \right.} \right) \times }  \\ 
 p\left( {{y_{u,m,i}}\left| {\widehat{\bf{r}}_u^{\left( k \right)},\hat \Omega _u^{\left( {k,z - 1} \right)},{C_u}} \right.} \right) \\ 
 {\rm{ = }} - \sum\limits_{{y_{u,m,i}}} {{{\left\| {\hat r_{u,m,i}^{\left( k \right)} - {e^{j{\theta _{u,m}}}}{e^{j{{2\pi {\varepsilon _u}\left( {i - 1} \right)} \mathord{\left/
 {\vphantom {{2\pi {\varepsilon _u}\left( {i - 1} \right)} N}} \right.
 \kern-\nulldelimiterspace} N}}}{y_{u,m,i}}} \right\|}^2} \times }  \\ 
 p\left( {{y_{u,m,i}}\left| {\widehat{\bf{r}}_u^{\left( k \right)},\hat \Omega _u^{\left( {k,z - 1} \right)},{C_u}} \right.} \right) \\ 
  =  - \sum\limits_{{y_{u,m,i}}} {\left\{ {{{\left\| {\hat r_{u,m,i}^{\left( k \right)}} \right\|}^2} + {{\left\| {{y_{u,m,i}}} \right\|}^2} - } \right.}  \\ 
 \left. { - 2Re\left\{ {{{\left( {\hat r_{u,m,i}^{\left( k \right)}} \right)}^H}{e^{j{\theta _{u,m}}}}{e^{j{{2\pi {\varepsilon _u}\left( {i - 1} \right)} \mathord{\left/
 {\vphantom {{2\pi {\varepsilon _u}\left( {i - 1} \right)} N}} \right.
 \kern-\nulldelimiterspace} N}}}{y_{u,m,i}}} \right\}} \right\} \times  \\ 
 p\left( {{y_{u,m,i}}\left| {\widehat{\bf{r}}_u^{\left( k \right)},\hat \Omega _u^{\left( {k,z - 1} \right)},{C_u}} \right.} \right) \\ 
 \end{array}
\end{equation}
where we have already dropped the constant term. Then, we can easily compute the following expectations
\begin{equation}\label{A2}
\begin{array}{l}
\sum\limits_{{y_{u,m,i}}} {{{\left\| {\widehat r_{u,m,i}^{\left( k \right)}} \right\|}^2}p\left( {{y_{u,m,i}}\left| {\widehat{\bf{r}}_u^{\left( k \right)},\widehat\Omega _u^{\left( {k,z - 1} \right)},{C_u}} \right.} \right)}  
= {\left\| {\widehat r_{u,m,i}^{\left( k \right)}} \right\|^2} \\ 
\sum\limits_{{y_{u,m,i}}} {{\mathop{\rm Re}\nolimits} \left\{ {{{\left( {\widehat r_{u,m,i}^{\left( k \right)}} \right)}^H}{e^{j{\theta _{u,m}}}}{e^{{{j2\pi {\varepsilon _u}\left( {i - 1} \right)} \mathord{\left/
						{\vphantom {{j2\pi {\varepsilon _u}\left( {i - 1} \right)} N}} \right.
						\kern-\nulldelimiterspace} N}}}{y_{u,m,i}}} \right\}}  \times  \\ 
p\left( {{y_{u,m,i}}\left| {\widehat{\bf{r}}_u^{\left( k \right)},\widehat\Omega _u^{\left( {k,z - 1} \right)},{C_u}} \right.} \right) \\ 
= Re\left\{ {{{\left( {\widehat r_{u,m,i}^{\left( k \right)}} \right)}^H}{e^{j{\theta _{u,m}}}}{e^{{{j2\pi {\varepsilon _u}\left( {i - 1} \right)} \mathord{\left/
					{\vphantom {{j2\pi {\varepsilon _u}\left( {i - 1} \right)} N}} \right.
					\kern-\nulldelimiterspace} N}}}{m_{{y_{u,m,i}}}}} \right\} \\ 
\sum\limits_{{y_{u,m,i}}} {{{\left\| {{y_{u,m,i}}} \right\|}^2}p\left( {{y_{u,m,i}}\left| {\widehat{\bf{r}}_u^{\left( k \right)},\widehat\Omega _u^{\left( {k,z - 1} \right)},{C_u}} \right.} \right)}  \\ 
= {\left\| {{m_{{y_{u,m,i}}}}} \right\|^2} + \sigma _{{y_{u,m,i}}}^2 \\ 
\end{array}
\end{equation}
where the mean and variance of ${y_{u,m,i}}$ are  given in the Gaussian expression for $p\left( {{y_{u,m,i}}\left| {\widehat{\bf{r}}_u^{\left( k \right)},\widehat\Omega _u^{\left( {k,z - 1} \right)},{C_u}} \right.} \right)$ in (\ref{gauss}). Substituting (\ref{A2}) in to (\ref{A1}), we have 
\begin{equation}
\begin{array}{l}
{Q_{m,i}}\left( {{\Omega _u}\left| {\widehat\Omega _u^{\left( {k,z - 1} \right)}} \right.} \right) \\ 
=  - {\left\| {\widehat r_{u,m,i}^{\left( k \right)}} \right\|^2} - {\left\| {{m_{{y_{u,m,i}}}}} \right\|^2} \\ 
+ 2{\mathop{\rm Re}\nolimits} \left\{ {{{\left( {\widehat r_{u,m,i}^{\left( k \right)}} \right)}^H}{e^{j{\theta _{u,m}}}}{e^{{{j2\pi {\varepsilon _u}i} \mathord{\left/
					{\vphantom {{j2\pi {\varepsilon _u}i} N}} \right.
					\kern-\nulldelimiterspace} N}}}{m_{{y_{u,m,i}}}}} \right\} - \sigma _{{y_{u,m,i}}}^2 \\ 
=  - {\left\| {\widehat r_{u,m,i}^{\left( k \right)} - {e^{j{\theta _{u,m}}}}{e^{{{j2\pi {\varepsilon _u}i} \mathord{\left/
						{\vphantom {{j2\pi {\varepsilon _u}i} N}} \right.
						\kern-\nulldelimiterspace} N}}}{m_{{y_{u,m,i}}}}} \right\|^2} - \sigma _{{y_{u,m,i}}}^2 \\ 
\end{array}
\end{equation}
Since  $\sigma _{{y_{u,m,i}}}^2$ is not relevant to ${\Omega _u}$. We just drop it. This gives the ultimate form of Q function in (\ref{sqf2}).

\subsection{Derivation of (\ref{cfoest})}

We denote the objective function in (\ref{cfoe}) by 
\begin{equation}\label{B1}
\begin{array}{l}
f\left( {{\varepsilon _u}} \right) \buildrel \Delta \over =  \\ 
- \sum\limits_{m = 1}^M {{{\left\| {\widehat{\bf{r}}_{u,m}^{\left( k \right)} - {e^{j\widehat\theta _{u,m}^{\left( {k,z - 1} \right)}}}{\bf{\Gamma }}\left( {{\varepsilon _u}} \right){{\bf{F}}^H}{\bf{D}}\left( {{{\bf{m}}_{{{\bf{X}}_{u,m}}}}} \right){\bf{F}}\widehat{\bf{h}}_u^{\left( {k,z - 1} \right)}} \right\|}^2}}  \\ 
\end{array}
\end{equation}
Expanding  and dropping the terms irrelevant to $\varepsilon _u$,  we obtain 
\begin{equation}\label{B2}
\begin{array}{l}
f\left( {{\varepsilon _u}} \right) = \sum\limits_{m = 1}^M {}  \\ 
{\mathop{\rm Re}\nolimits} \left\{ {{{\left( {\widehat{\bf{r}}_{u,m}^{\left( k \right)}} \right)}^H}{e^{j\widehat\theta _{u,m}^{\left( {k,z - 1} \right)}}}{\bf{\Gamma }}\left( {{\varepsilon _u}} \right){{\bf{F}}^H}{\bf{D}}\left( {{{\bf{m}}_{{{\bf{X}}_{u,m}}}}} \right){\bf{F}}\widehat{\bf{h}}_u^{\left( {k,z - 1} \right)}} \right\} \\ 
\end{array}
\end{equation}
We choose to approximate ${\bf{\Gamma }}\left( {{\varepsilon _u}} \right)$ in the objective function using its Taylor series expansion 
\begin{equation}\label{B3}
\begin{array}{l}
{\bf{\Gamma }}\left( {{\varepsilon _u}} \right) \approx {\bf{\Gamma }}\left( {\widehat\varepsilon _u^{\left( {k,z - 1} \right)}} \right) + \left( {{\varepsilon _u} - \widehat\varepsilon _u^{\left( {k,z - 1} \right)}} \right){\bf{\Gamma '}}\left( {\widehat\varepsilon _u^{\left( {k,z - 1} \right)}} \right) \\ 
+ \frac{1}{2}{\left( {{\varepsilon _u} - \widehat\varepsilon _u^{\left( {k,z - 1} \right)}} \right)^2}{\bf{\Gamma ''}}\left( {\widehat\varepsilon _u^{\left( {k,z - 1} \right)}} \right) \\ 
\end{array}
\end{equation}
where we truncate the third-order terms and above and ${\widehat\varepsilon _u^{\left( {k,z - 1} \right)}}$ is the starting point. Substituting (\ref{B3}) into (\ref{B2}) and differentiating the resulting $f\left( {{\varepsilon _u}} \right)$
 with respect to ${{\varepsilon _u}}$ yield
$$\begin{array}{l}
\frac{{\partial f\left( {{\varepsilon _u}} \right)}}{{\partial {\varepsilon _u}}} = \sum\limits_{m = 1}^M {\left\{ {{{\left( {\widehat{\bf{r}}_{u,m}^{\left( k \right)}} \right)}^H}{e^{j\hat \theta _{u,m}^{\left( {k,z - 1} \right)}}}} \right.\left[ {\Gamma '\left( {\hat \varepsilon _{\bf{u}}^{\left( {{\bf{k}},{\bf{z}} - {\bf{1}}} \right)}} \right)} \right.}  \\ 
+ \left. {\left( {{\varepsilon _u} - \hat \varepsilon _u^{\left( {k,z - 1} \right)}} \right)\Gamma ''\left( {\hat \varepsilon _{\bf{u}}^{\left( {{\bf{k}},{\bf{z}} - {\bf{1}}} \right)}} \right)} \right] \times  \\ 
\left. {{{\bf{F}}^H}{\bf{D}}\left( {{{\bf{m}}_{{{\bf{X}}_{u,m}}}}} \right){\bf{F}}\widehat{\bf{h}}_u^{\left( {k,z - 1} \right)}} \right\} \\ 
\end{array}$$
Finally, setting  ${{\partial f\left( {{\varepsilon _u}} \right)} \mathord{\left/
		{\vphantom {{\partial f\left( {{\varepsilon _u}} \right)} {\partial {\varepsilon _u}}}} \right.
		\kern-\nulldelimiterspace} {\partial {\varepsilon _u}}} = 0$ and solving the equation, we obtain the new CFO update shown in (\ref{cfoest}). 

\subsection{Discussion of Receiver in \cite{pun2007iterative} }

Ref. \cite{pun2007iterative} proposed an iterative receiver for solving the problem of joint channel-parameter estimation, CFO compensation and data detection in uncoded OFDMA systems. The authors of \cite{pun2007iterative} attempted to construct the receiver based on a SAGE-ECM framework. As explained later in this appendix, this attempt ended up with a pure SAGE framework due to an approximation.

A major difference between the framework in \cite{pun2007iterative} and ours is the assignment of hidden data in the single-user subproblem within the SAGE-ECM framework. For us, the data symbols ${{\bf{X}}_{u}}$ are the hidden data and the channel gains ${{\bf{h}}_u}$ are one of the parameters. For in \cite{pun2007iterative}, the channel gains ${{\bf{h}}_u}$ are the hidden data, and the data symbols ${{\bf{X}}_{u}}$ are one of the parameters.

Another difference is that, unlike our work here, \cite{pun2007iterative} did not incorporate channel coding/decoding --- \cite{pun2007iterative} only focused on detection of individual uncoded OFDM blocks ${{\bf{X}}_{u,m}}$, and the correlations among different data blocks of the overall frame ${{\bf{X}}_{u}}$ induced by channel coding are not exploited to further improve performance. In particular, each OFDM symbols are decoded independently. Here, for consistency with our problem, we extend the treatment of \cite{pun2007iterative} to coded OFDM-IDMA systems  (i.e., we incorporate channel coding into the framework of \cite{pun2007iterative}). With this extension, we have ${\Omega '_u} = \left\{ {{\varepsilon _u},\left\{ {{\theta _{u,m}}} \right\}_{m = 1}^M,{{\bf{X}}_u}} \right\}$ as the parameters and ${{\bf{h}}_u}$ as the hidden data. 

The corresponding ECM aims to solve the following problem iteratively: 
\begin{equation}\label{ecm22}
\begin{array}{l}
\left( {\widehat\varepsilon _u^{\left( k \right)},\left\{ {\widehat\theta _{u,m}^{\left( k \right)}} \right\}_{m = 1}^M,\widehat{\bf{X}}_u^{\left( k \right)}} \right) \\ 
= \arg \mathop {\max }\limits_{\left( {{\varepsilon _u},\left\{ {{\theta _{u,m}}} \right\}_{m = 1}^M,{{\bf{X}}_u} \in {C_u}} \right)} \log p\left( {\widehat{\bf{r}}_u^{\left( k \right)}\left| {{\varepsilon _u},\left\{ {{\theta _{u,m}}} \right\}_{m = 1}^M,{{\bf{X}}_u}} \right.} \right) \\ 
= \arg \mathop {\max }\limits_{\left( {{\varepsilon _u},\left\{ {{\theta _{u,m}}} \right\}_{m = 1}^M,{{\bf{X}}_u} \in {C_u}} \right)}  \\ 
\;\;\;\;\;\;\;\;\;\;\;\;\;\;\;\;\;\;  \left\{ {\log \int {p\left( {\widehat{\bf{r}}_u^{\left( k \right)},{{\bf{h}}_u}\left| {{\varepsilon _u},\left\{ {{\theta _{u,m}}} \right\}_{m = 1}^M,{{\bf{X}}_u}} \right.} \right)d} {{\bf{h}}_u}} \right\} \\ 
\end{array}
\end{equation}


Now, ECM solves (\ref{ecm22}) iteratively. In the $z^{th}$ iteration of ECM, the E-step of ECM computes the Q function of  
\begin{equation}\label{q_ecm2}
	\begin{array}{l}
		Q\left( {{{\Omega '}_u}\left| {\widehat{\Omega '}_u^{\left( {k,z - 1} \right)}} \right.} \right) \\ 
		= {E_{{{\bf{h}}_u}}}\left\{ {\ln p\left( {\widehat{\bf{r}}_{u,m}^{\left( k \right)}\left| {{{\bf{h}}_u},{{\Omega '}_u}} \right.} \right)p\left( {\widehat{\bf{r}}_{u,m}^{\left( k \right)}\left| {{{\bf{h}}_u},\widehat{\Omega '}_u^{\left( {k,z - 1} \right)}} \right.} \right)} \right\} \\ 
	\end{array}
\end{equation}
and the M-step of ECM updates the tentative estimates $\widehat{\Omega '}_u^{\left( {k,z - 1} \right)}$ for  ${\Omega '_u}$ by finding the  ${\Omega '_u}$  that maximize (\ref{ecm22}). 

The exact computation of the Q function in (\ref{q_ecm2}) is complex and not feasible from the implementation viewpoint  (cf .eq (13) in \cite{pun2007iterative} for the Q function of the uncoded case). To reduce complexity, \cite{pun2007iterative} made an approximation on the computed Q function. Extending the result in \cite{pun2007iterative} to channel-coded case (cf. eq (16) in \cite{pun2007iterative}), the E-step of ECM approximates the Q function as
\begin{equation}\label{ecm2_q_a}
	\begin{array}{l}
		Q\left( {{{\Omega '}_u}\left| {\widehat{\Omega '}_u^{\left( {k,z - 1} \right)}} \right.} \right) \\ 
		=  - \sum\limits_{m = 1}^M {\left\| {\widehat{\bf{r}}_{u,m}^{\left( k \right)} - {e^{j{\theta _{u,m}}}}{\bf{\Gamma }}\left( {{\varepsilon _u}} \right){{\bf{F}}^H}{\bf{D}}\left( {{{\bf{X}}_{u,m}}} \right)} \right.}  \\ 
		\;\;\;\;\;\;\;\;\;\;\;\;\;\;\;\;\;\;\;\;\;\;\;\;\;\;\;\;\;\;\;\;\;\;\;\;\;\;\;\;\;\;\;\;\;\;\;\;\;\;\;\;\;  {\left. { \times {\bf{F}}\widehat{\bf{h}}_u^{\left( {LS} \right)}\left( {{{\widehat\Omega } }_u^{\prime \left( {k,z - 1} \right)}} \right)} \right\|^2} \\ 
	\end{array}
\end{equation} 
where $\widehat{\bf{h}}_u^{\left( {LS} \right)}\left( {{{\widehat\Omega }}_u^{\prime  \left( {k,z - 1} \right)}} \right)$ is the LS estimate of ${{\bf{h}}_u}$ obtained with ${\Omega _u}^{\prime} = {\widehat\Omega  }_u^{\prime \left( {k,z - 1} \right)}$ in the signal model (\ref{sm2}).  With this Q function, the CFO and the phases can be updated to $\widehat\varepsilon _u^{\left( {k,z} \right)}$ and $\left\{ {\theta _{u,m}^{\left( {k,z} \right)}} \right\}_{m = 1}^M$ using the same methods of (\ref{cfoest}) and (\ref{phaest}) in Section III.C, respectively. However, the transmit symbols  ${{\bf{X}}_u}$  here is updated in a different way than in Section III.C. Specifically, the stage in the M-step for updating the transmit symbols now becomes (\ref{ecm23}), which can be solved using the min-sum channel decoding, as explained in Section III.D.

A subtle but important consequence of approximating the Q function as in (\ref{ecm2_q_a}) is that, although the development of the receiver in \cite{pun2007iterative} begins with the ECM algorithm, it ends up with a pure SAGE solution to (\ref{subprob}). In other words, it is not a SAGE-ECM algorithm anymore, but a pure SAGE algorithm. The algorithm essentially treats all the variables in $\left\{ {{\varepsilon _u},\left\{ {{\theta _{u,m}}} \right\}_{m = 1}^M,{{\bf{h}}_u},{{\bf{X}}_u}} \right\}$ as parameters, and none of them as hidden data. As in the SAGE algorithm, the parameters are estimated sequentially one at a time; when one parameter is under estimation, all other parameters are fixed to their estimates from the last iteration. This pure SAGE algorithm is exactly the same as that we discussed in Section III.D and adopted as one of the benchmarks in our simulation studies in Section IV. 

\ifCLASSOPTIONcaptionsoff
  \newpage
\fi

\bibliographystyle{IEEEtran}
\bibliography{database}


\end{document}